\documentclass[nohyper,12pt,letterpaper]{JHEP3}
\usepackage{epsfig,youngtab}

\newcommand{\myfig}[3]{\begin{figure}[ht]
\begin{center}
\leavevmode \epsfxsize=#2cm \epsfbox{#1}
\end{center}
\caption{#3} \label{fig:#1}
\end{figure}}

\author{Robert de Mello Koch$^{1,2}$\\
\qquad \\
$^{1}$ National Institute for Theoretical Physics,\\
Department of Physics and Centre for Theoretical Physics,\\ 
University of the Witwatersrand,\\ 
Wits, 2050,\\ 
South Africa\\
\qquad\\
$^{2}$Stellenbosch Institute for Advanced Studies,\\
Stellenbosch,\\
South Africa\\
\qquad\\
E-mail: \email{robert@neo.phys.wits.ac.za}}

\abstract{
Type IIB string theory on spacetimes that are asymptotically AdS$_5\times$S$^5$ can be defined using four dimensional ${\cal N}=4$ super 
Yang-Mills theory. Six of the dimensions of the string theory are holographically reconstructed in the Yang-Mills theory. In this article
we study how these dimensions and local physics in these dimensions emerge. We reorganize the dynamics of the ${1\over 2}$ BPS sector of
the field theory by rewriting it in terms of Schur polynomials. The Young diagram labeling of these polynomials can be viewed as a book
keeping device which summarizes how the operator is constructed. We show that aspects of the geometry of the extra holographic dimensions
are captured very naturally by the Young diagram. Gravitons which are localized at a specific position in the bulk correspond to boxes added 
at a specific location on the Young diagram. 
}

\preprint{WITS-CTP-038}

\title{Geometries from Young Diagrams}


\def \Tr{\mbox{Tr\,}}

\begin{document}

\section{Introduction}

We do not yet know what string theory is. AdS/CFT\cite{Maldacena:1997re} seems to provide us with a definition of string theory
on negatively curved spaces; for example, type IIB string theory on spacetimes that are asymptotically 
AdS$_5\times$S$^5$ is given by four dimensional ${\cal N}=4$ super Yang-Mills theory (see \cite{Lin:2004nb} for an explicit
demonstration of this in the ${1\over 2}$-BPS sector).
There are many aspects of this definition that need to be clarified. The string theory is a 10 dimensional theory; apparently, there are
an additional 6 dimensions that are holographically reconstructed in the Yang-Mills theory. Understanding how these dimensions emerge 
and further, how {\it local physics} in these dimensions emerge is a fascinating problem. In this article, we explore this issue. 

The dynamical content of any quantum field theory is captured in its correlation functions. The extra 
holographic dimensions together with their local dynamics should be coded into the correlation functions of ${\cal N}=4$ super 
Yang-Mills theory. See \cite{Skenderis:2007yb} for detailed examples of what information is coded and how its coded. 
According to the AdS/CFT correspondence, we expect a 
classical geometry to emerge in the strong 't Hooft coupling limit of the field theory. In this article, we focus mainly on correlation
functions of ${1\over 2}$ BPS operators because these correlators are protected\cite{Lee:1998bxa}. 
Thus, we will compute these correlation functions at weak coupling with confidence
that these free field results can be extrapolated, with no change, to strong coupling.

Our approach entails performing a reorganization of the dynamics of the field theory. The ${1\over 2}$ BPS sector of the theory is captured by 
the singlet dynamics of a single holomorphic matrix\cite{Corley:2001zk,Berenstein:2004kk}. 
The reorganization we employ is achieved by a change of variables: we will use Schur polynomials which provide a complete description of 
the ${1\over 2}$ BPS sector\cite{Corley:2001zk}\footnote{They are simply an alternative basis to the trace basis.}. 
This reorganization will allow a significant simplification of the computation
of correlators for two reasons. First, the Schur polynomials satisfy a nice product rule that can be used to collapse any product of
Schur polynomials into a sum of Schur polynomials. Second, the two point functions of Schur polynomials, in the free field theory limit,
are known exactly\cite{Corley:2001zk}. Thus, this reorganization provides a considerable simplification of correlator 
calculations. The label of the Schur polynomial, a Young diagram,
plays a central role.

We will start with a quick review of Young diagrams. We have two goals in mind: (i) to recall the definition of the hooks and
the weights of a Young diagram and (ii) to stress that the hooks and weights encode the specific way in which a tensor has
been symmetrized or antisymmetrized.

Young diagrams are a useful way to label the irreducible representations of $SU(N)$\footnote{A Young diagram with $n$ boxes also labels an
irreducible representation of the symmetric group $S_n$. We will also exploit this connection in what follows.}. A tensor transforming in an irreducible
representation of $SU(N)$ has a definite symmetry under permutations of its indices. It is this symmetry that the Young diagram
records. A Young diagram with a single column containing $n$-boxes corresponds to a completely antisymmetric tensor with $n$ 
indices. A Young diagram with a single row containing $m$-boxes corresponds to a completely symmetric tensor with $m$ indices. 
To compute the dimension of an $SU(N)$ irreducible representation corresponding to a particular Young diagram, we need to define 
the weight\footnote{These are not the Dynkin weights.} and the hook of each box in the Young diagram. A box in row $i$ and column 
$j$ has a weight equal to $N-i+j$. Here is an example of a Young diagram with the weights filled in

\vfill\eject

\myfig{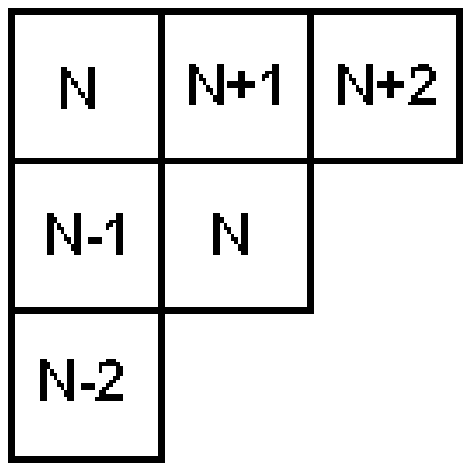}{3.0}{A Young diagram with the weight of each box displayed.}

To obtain the hook associated to a given box, draw a line starting from the given box towards the bottom of
the page until you exit the Young diagram, and another line starting from the same box towards the right until
you again exit the diagram. These two lines form an elbow - what we call the hook. An example of a hook is 
\myfig{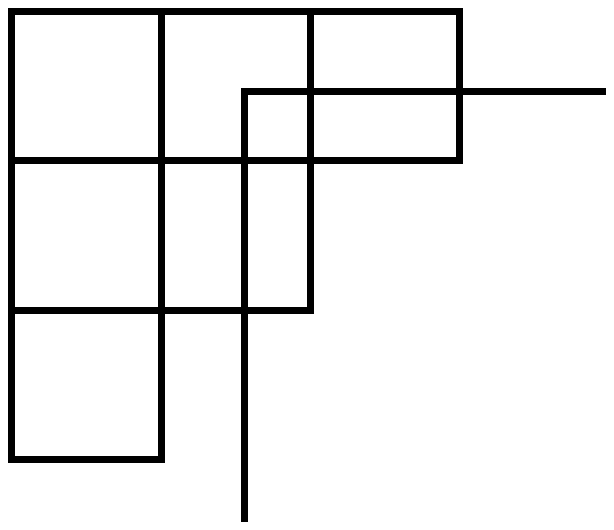}{3.0}{The hook shown is associated with the box in which the corner of the elbow lies.}

The hook length for the given box is obtained by counting the number of boxes the elbow belonging to the box passes through. Here 
is a Young diagram with the hook lengths filled in
\myfig{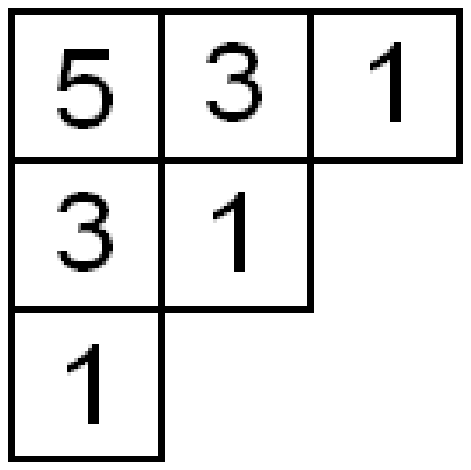}{2.6}{A Young diagram with the hook length of each box displayed.}

The dimension of the $SU(N)$ irreducible representation corresponding to this Young diagram is given by the product of the
weights divided by the product of the hooks. For the example we are considering
$$ d_{\tiny \yng(3,2,1)}={N^2 (N^2-1)(N^2-4)\over 5\cdot 3\cdot 3}. $$
The hooks and the weights provide an efficient way to encode the combinatorics of tensors with a definite symmetry under swapping 
indices. For example, the totally antisymmetric tensor with three indices, gives a non-zero result only if all indices take distinct
values. Thus, the first index can take any one of $N$ values, the second index any one of $N-1$ values and the third index any one
of $N-2$ values. These are exactly the value of the weights of the corresponding Young diagram. The division by the hooks corrects 
for the fact that not all elements of this tensor are independent - swapping any two indices only costs a sign. 

Recently is has become clear that Young diagrams also provide a useful labeling of the invariant variables of matrix models.
The first such example was given in \cite{Corley:2001zk}, where it was shown that Schur polynomials built using a single complex
scalar $Z$ provide a convenient parametrization of the ${1\over 2}$ BPS sector of the theory. A particularly nice property of the 
Schur polynomials 
$$\chi_R(Z)={1\over n!}\sum_{\sigma\in S_n}\chi_R(\sigma )Z^{i_1}_{i_{\sigma (1)}}\cdots Z^{i_n}_{i_{\sigma (n)}},$$
is that they have a diagonal two point function
\begin{equation}
\left\langle\chi_{R}(Z)^\dagger\chi_S(Z)\right\rangle = f_R\delta_{RS}\, .
\label{SchurTwoPoint}
\end{equation}
If $R$ is a Young diagram with $n$ boxes, then the Schur polynomial is built using products of traces of Higgs fields $Z$ such that 
each term is a product of $n$ $Z$s. The factor $f_R$ is equal to the product of the weights of $R$. In addition to this, the Schur polynomials
also satisfy a product rule called the Littlewood-Richardson rule 
\begin{equation}
\chi_{R_1}(Z)\chi_{R_2}(Z) = \sum_{S}g_{R_1\, R_2\, S}\chi_S(Z)\, .
\label{LRrule}
\end{equation}
The Littlewood-Richardson number $g_{R_1\, R_2\, S}$ counts the number of times the $SU(N)$ irreducible representation $S$ appears in the tensor product
$R_1\otimes R_2$. The product rule follows immediately from the fact that when evaluated on a unitary matrix $U$, $\chi_R(U)$ gives
the character of $U$ in irreducible representation $R$.
These results have been extended to define operators in multimatrix models, that have diagonal two point functions and whose multipoint
functions again follow from group theory reasoning\cite{Brown:2007xh,Kimura:2007wy,Bhattacharyya:2008rb}.
For a recent review see \cite{sanjaye}.

According to the AdS/CFT correspondence\cite{Maldacena:1997re}, these operators have a description in a dual quantum gravity. In particular,
for operators with an ${\cal R}$-charge of order $N^2$ the Schur polynomials should be dual to new geometries\cite{Lin:2004nb}. 
We can now sharpen the question we are asking: the goal of 
this article is to investigate how this dual geometry emerges. If the Young diagram labeling of operators is to provide a useful tool to
explore the AdS/CFT correspondence, the dual geometry should emerge naturally from the Young diagram. We will argue that this is indeed
the case. Our approach to this problem entails defining coherent states in the gauge theory that we propose are dual to graviton states. This
gives us a way to explore the dual geometry: gravitons move along null geodesics and hence they ``know'' about the dual geometry.
To define the coherent states, we need to define graviton creation and annihilation operators in the Yang-Mills theory. 
The graviton annihilation and creation operators that we define add or remove boxes to the Young diagram describing the background.
The state with a box added/removed needs to be normalized by a factor of $\sqrt{c\over N}$ where $c$ is the weight of the added/removed box.
The extra $\sqrt{c\over N}$ factor plays a key role in ensuring that the correct dual geometry emerges.

In the next section we start by discussing gravitons in the AdS$_5\times$S$^5$ background. This serves to illustrate the main ideas in the
simplest possible setting. In section 3 we turn to consider LLM geometries; section 4 is concerned with geometries that arise when the field
theory background is built from two matrices. In section 5 we discuss our results.

To conclude this introduction, we mention the papers \cite{Vazquez:2006id,Chen:2007gh}. By probing the ${1\over 2}$ BPS operators in the 
gauge theory with single trace operators that correspond to closed strings these papers have already understood some aspects of the geometry
of the dual IIB supergravity. These articles used an eigenvalue density description for the ${1\over 2}$ BPS operators in the gauge theory.
Using the results of section 3.3 it is easy to see that where our results and those of \cite{Vazquez:2006id,Chen:2007gh} overlap, they 
are consistent.

\section{Gravitons in AdS$_5\times$S$^5$}

${\cal N}=4$ super Yang-Mills theory is a conformal field theory. By the state/operator correspondence, the theory on 4 dimensional Euclidean 
space can be mapped to a theory on $R\times S^3$. Under this map operators map to states and conformal dimensions of operators map into
energies of states. We will primarily be interested in the half-BPS operators that can be constructed from the s-waves of complex combinations
of the adjoint scalars of the theory on $R\times S^3$. We will make use of two scalars $Z,Y$; they have two point function (we suppress the
spacetime dependence of this correlator as these spacetime dependences will play no role in this article)
$$\left\langle Z_{ij}Z^\dagger_{kl}\right\rangle =\delta_{il}\delta_{jk}=\left\langle Y_{ij}Y^\dagger_{kl}\right\rangle \, .$$
Our computations will mostly be performed in the theory on four dimensional Euclidean space, where we can use the known results for 
correlators of Schur polynomials and restricted Schur polynomials. To make contact with the dual gravity we will conformally map our
results to the theory on $R\times S^3$. The boundary of AdS$_5\times$S$^5$ in global coordinates is $R\times S^3$ - it is the
space that the field theory is defined on; this is why it is easiest to interpret the field theory after mapping to $R\times S^3$.
For the operators we consider, after mapping to $R\times S^3$, the dynamics reduces to that of $N$ non-relativistic fermions
in an external potential\cite{Corley:2001zk,Berenstein:2004kk,Brezin:1977sv}. The phase space 
of these fermions can be identified with the moduli space of the dual geometries\cite{Lin:2004nb}.

Chiral primary operators in ${\cal N}=4$ super Yang-Mills theory, with ${\cal R}$ charge of $O(1)$ are dual to Kaluza-Klein gravitons
on the AdS$_5\times$S$^5$ background\cite{Maldacena:1997re}. 
The ${\cal R}$-charge of the operator maps into the angular momentum (on the $S^5$) of the graviton.
The operator dual to a graviton with one unit of angular momentum is
$$ {\Tr (Z)\over\sqrt{N}}.$$
The factor of $N^{-{1\over 2}}$ ensures that this operator has a unit two point function and is hence dual to a normalized state. 
It is straight forward to compute
\begin{equation}
\left\langle \left({\Tr (Z^\dagger )\over\sqrt{N}}\right)^n\left({\Tr (Z)\over\sqrt{N}}\right)^m \right\rangle =n!\delta_{m,n}\, .
\label{gravitoncorr}
\end{equation}
This suggests that the action of $\Tr (Z)/\sqrt{N}$ on the vacuum in the super Yang-Mills theory matches the action of the graviton creation
operator in the dual quantum gravity on AdS$_5\times$S$^5$. We will write connections like this using a double sided arrow 
$$ {\Tr (Z)\over\sqrt{N}}\leftrightarrow a^\dagger\, ,$$
where $a^\dagger$ is a graviton creation operator. It is useful
to use the coordinates of \cite{Lin:2004nb}. In these coordinates, there is an $S^3$ extracted from the $S^5$. The remaining two coordinates
of the $S^5$ is combined with the radial direction of AdS$_5$ to make three new coordinates, $y,x^1,x^2$. We can also describe the $x^1,x^2$ plane
using an angle $\phi$ and a radius $r$. The graviton that is created is an $s$-wave on the field theory $S^3$.
It is also an $s$-wave on the second $S^3$ (extracted from $S^5$) and is smeared along the $\phi$ direction. It is
localized at $y=0$ and at some fixed $r$. It is this localization that we are trying to describe. A normalized $n$-graviton state is dual to the operator
$$ {1\over\sqrt{n!}}\left({\Tr (Z)\over\sqrt{N}}\right)^n \, .$$
We can also define the graviton annihilation operator as
$${1\over\sqrt{N}}\Tr \left({d\over dZ}\right)\leftrightarrow a\, .$$
Note that
\begin{equation}
\big[ a,a^\dagger \big]=1\leftrightarrow \big[ {1\over\sqrt{N}}\Tr \left({d\over dZ}\right),{\Tr (Z)\over\sqrt{N}} \big]=
{1\over N}\left(\Tr {d\over dZ} \, \Tr Z\right) =1.
\label{comm}
\end{equation}

As a check of these identifications, we will now study the dynamics of coherent states of gravitons and verify that we recover the
expected graviton dynamics. Define an operator dual to a graviton coherent state as
$$ {\cal O}_z = {\cal N}\exp ({\Tr (Z)\over\sqrt{N}} z),\qquad \big[ {1\over\sqrt{N}}\Tr \left({d\over dZ}\right),{\cal O}_z\big] = z{\cal O}_z \, .$$  
The normalization ${\cal N}$ is easily determined ($\bar{z}$ is the complex conjugate of $z$)
$$ \left\langle {\cal O}_z^\dagger {\cal O}_z\right\rangle = 
|{\cal N}|^2\sum_{n=0}^\infty\sum_{m=0}^\infty \left\langle \left({\Tr(Z)\over\sqrt{N}} \right)^n \left({\Tr (Z)^\dagger\over\sqrt{N}} \right)^m \right\rangle 
{z^n\over n!}{\bar{z}^m\over m!}=|{\cal N}|^2\sum_{n=0}^\infty {|z|^{2n}\over n!}=|{\cal N}|^2 e^{|z|^2}=1,$$
$$ {\cal O}_z =e^{-{1\over 2}|z|^2}e^{z{\Tr (Z)\over\sqrt{N}}} \, .$$
Since we study ${1\over 2}$ BPS operators, the conformal dimension of our operators (which is the graviton's energy in the dual description)
equals their ${\cal R}$-charge
$$ \big[ {\cal R},{\cal O}_z\big]=e^{-{1\over 2}|z|^2}\sum_{n=0}^\infty n {z^n \Tr (Z)^n\over n!\sqrt{N^n}}.$$ 
After conformally mapping to $R\times S^3$, the operator ${\cal O}_z$ is mapped to the coherent state $|z\rangle$, 
so that ($z=re^{-i\phi}$)
$$ \left\langle{\cal O}_z^\dagger \big[ {\cal R},{\cal O}_z\big]\right\rangle = \langle z|H|z \rangle=\bar{z}z=r^2, \qquad
\left\langle{\cal O}_z^\dagger i{d\over dt} {\cal O}_z\right\rangle = \langle z|i{d\over dt}|z\rangle .$$ 
The Lagrangian governing the low energy excitations of this coherent state on $R\times S^3$ is
$$ L= \langle z|i{d\over dt}|z \rangle - \langle z|H|z \rangle = \dot{\phi} r^2 -r^2 .$$
The equations of motion determine $\dot{\phi}=1$ and $\dot{r}=0$, which are by now, familiar results for gravitons in 
AdS$_5\times$S$^5$\cite{McGreevy:2000cw}.

Before moving on we will give a discussion of (\ref{gravitoncorr}). This formula can be obtained using elementary (just Wick contract)
methods. We will give a derivation of it using Schur technology since this will generalize nicely in the next section. The operator $\Tr (Z)$ is 
equal to the Schur polynomial $\chi_{\tiny \yng(1)}(Z)$. A product of Schur polynomials is easily computed using the Littlewood-Richardson rule.
Using this rule, it is straight forward to verify that ($d_R$ is the dimension of the $S_n$ irreducible representation labeled by $R$)
$$ (\Tr (Z))^n=(\chi_{\tiny \yng(1)}(Z))^n=\sum_R d_R\chi_R(Z),$$
and hence that ($R$ is a Young diagram with $n$ boxes; $S$ is a Young diagram with $m$ boxes; the sums run over all possible diagrams with
the correct number of boxes)
\begin{eqnarray}
\left\langle \left({\Tr (Z^\dagger )\over\sqrt{N}}\right)^n\left({\Tr (Z)\over\sqrt{N}}\right)^m \right\rangle
&=&\left\langle \left({\chi_{\tiny \yng(1)}(Z)^\dagger \over\sqrt{N}}\right)^n\left({\chi_{\tiny \yng(1)}(Z)\over\sqrt{N}}\right)^m \right\rangle
\nonumber\\
&=&\delta_{m,n}\sum_{R}\sum_{S}{1\over N^n} d_R d_S \left\langle \chi_R(Z)^\dagger\chi_S (Z)\right\rangle
\nonumber\\
&=&\delta_{m,n}\sum_R (d_R)^2 {f_R\over N^n}\nonumber\\
&=&\delta_{m,n}\sum_R (d_R)^2 \left( 1+ O\left({1\over N^2}\right)\right)\nonumber\\
&=& \delta_{m,n} n!\left( 1+ O\left({1\over N^2}\right)\right)\, ,\nonumber
\end{eqnarray}
where we have used the fact that $\sum_R (d_R)^2$ is equal to the order of the group. From our more elementary
techniques we know that all corrections to this large $N$ result vanish. Note that this multipoint correlator
was computed using the product rule and the two point function for Schur polynomials.

Finally, consider
\begin{equation}
\left\langle\left({\Tr (Z^{p\, \dagger} )\over\sqrt{pN^p}}\right)^n\left({\Tr (Z^p)\over\sqrt{pN^p}}\right)^m\right\rangle =n!\delta_{m,n}
\left( 1+ O\left({1\over N^2}\right)\right)\, .
\label{gravitoncorrtwo}
\end{equation}
This suggests that
$$ {\Tr (Z^p)\over\sqrt{pN^p}} $$
creates gravitons which each carry $p$ units of angular momentum. The result (\ref{gravitoncorrtwo})
can again be obtained by using Schur technology.
To do so, we need an expression for $\Tr (Z^p)$, which has been given in \cite{Corley:2002mj}. For example, for $p=2,3,4$ we have
(in general the right hand side is a sum over all hooks with an alternating sign; the first term corresponding to the 
symmetric representation is always positive)
$$ \Tr (Z^2)=\chi_{\tiny \yng(2)}(Z)-\chi_{\tiny \yng(1,1)}(Z),$$
$$ \Tr (Z^3)=\chi_{\tiny \yng(3)}(Z)-\chi_{\tiny \yng(2,1)}(Z)+\chi_{\tiny \yng(1,1,1)}(Z),$$
$$ \Tr (Z^4)=\chi_{\tiny \yng(4)}(Z)-\chi_{\tiny \yng(3,1)}(Z)+\chi_{\tiny \yng(2,1,1)}(Z)-\chi_{\tiny \yng(1,1,1,1)}.$$

\section{LLM Geometries}

In the previous section, we applied $\Tr Z$ and $\Tr {d\over dZ}$ to the vacuum of the super Yang-Mills theory to study gravitons in
AdS$_5\times$S$^5$. In this section we will study gravitons propagating on different backgrounds. These backgrounds correspond, in the
super Yang-Mills theory, to the operator $\chi_B (Z)$ where $B$ is a Young diagram with $O(N^2)$ boxes. 

\subsection{The Annulus}

A simple warm up example is provided by the case that $B$ is a Young diagram with $M$ columns
and each column contains exactly $N$ boxes; we take $M$ to be $O(N)$ so that $\mu\equiv {M\over N}$ is a number of $O(1)$. Let $R$ be
a Young diagram with $n=O(1)$ boxes. We will use the notation $+R$ to denote the Young diagram obtained by stacking $R$ next to B, and the
notation $-R$ to denote the Young diagram obtained by removing $R$ from the ``bottom right corner'' of $B$ as shown in figure 4 below.

\myfig{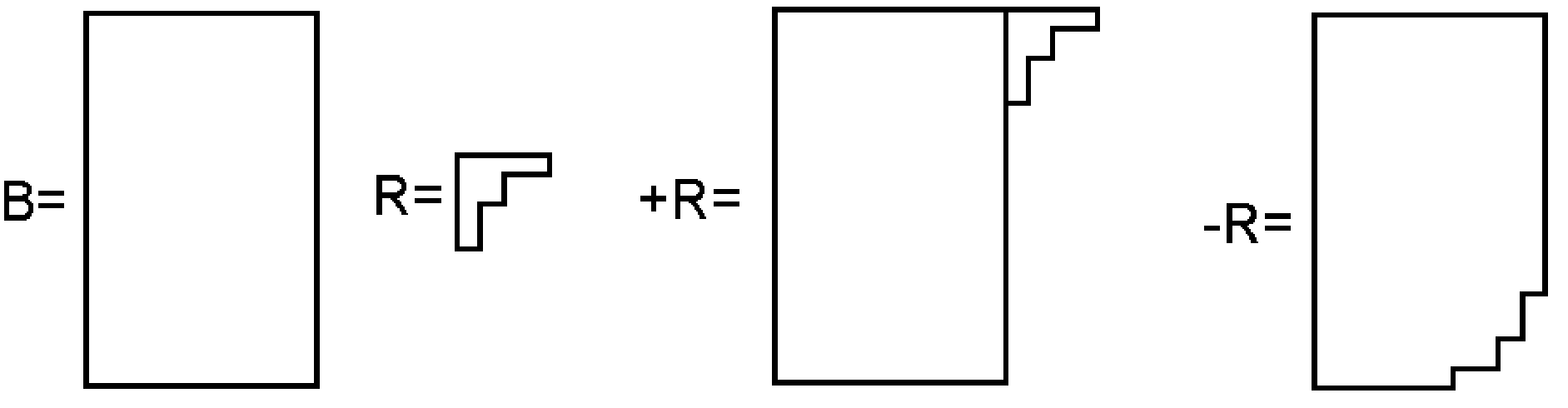}{15.0}{The definition of $+R$ and $-R$ in terms of $R$ and $B$.}

Our background is given by taking the ``vacuum'' to be the normalized state dual to the operator
$$|B \rangle \leftrightarrow {\chi_B(Z)\over\sqrt{f_B}} .$$
Gravitons on this background will correspond to (small) fluctuations about this state\footnote{By small we
mean small enough that backreaction can be neglected. This is the case since we take $n$ to be $O(1)$.}. 
An operator dual to an $n$ graviton state in the background $B$ is
$$ \left({\Tr (Z)\over\sqrt{N}}\right)^n {\chi_B(Z)\over\sqrt{f_B}}
=\left({\chi_{\tiny \yng(1)}(Z)\over\sqrt{N}}\right)^n {\chi_B(Z)\over\sqrt{f_B}}
=\sum_{R}d_R{\chi_{+ R} (Z)\over \sqrt{f_B N^n}} \, . $$
This formula is exact; to get this result, we used the product rule for Schur polynomials. 
We will now compute the two point correlator of this operator, to the leading order in the large $N$ limit
\begin{equation} 
\left\langle {\chi_B(Z)^\dagger\over\sqrt{f_B}}\left({\Tr (Z)^\dagger\over\sqrt{N}}\right)^n\left({\Tr (Z)\over\sqrt{N}}\right)^n {\chi_B(Z)\over\sqrt{f_B}}\right\rangle
=\sum_{R}(d_R)^2{f_{+ R}\over f_B N^n}=(1+\mu)^n\, n!\, . 
\label{corr1}
\end{equation}
To see the last equality note that all weights in $f_B$ cancel with weights in $f_{+ R}$, leaving only the weights corresponding to the boxes added to
$B$ to produce $+R$; these boxes all have a weight of $N+M+O(1)$ where the $O(1)$ number depends on exactly which box is considered.
Given this correlator, a normalized $n$-graviton state, to leading order in $N$, is dual to the operator
$$ {\cal O}_n ={1\over (1+\mu)^{n/2}\sqrt{n!}}\left({\chi_{\tiny \yng(1)}(Z)\over\sqrt{N}}\right)^n{\chi_B(Z)\over\sqrt{f_B}} .$$
Thus, in this new background
\begin{equation}
{\Tr (Z)\over\sqrt{N}}{\cal O}_n = \sqrt{c\over N} \sqrt{n+1}{\cal O}_{n+1} =\sqrt{1+\mu}\sqrt{n+1} {\cal O}_{n+1} \, ,
\label{zact}
\end{equation}
where $c$ is the weight of the box added to the background Young diagram $B$ after multiplying by $\Tr (Z)=\chi_{\tiny \yng(1)}(Z)$.
We can write 
$$ {\Tr (Z)\over\sqrt{N}}\leftrightarrow a^\dagger$$
with $a^\dagger$ a graviton creation operator in the background $B$. From (\ref{zact}) we can read off its action on an $n$ graviton state 
$$ a^\dagger |n\rangle = \sqrt{c\over N}\sqrt{n+1}|n+1\rangle = \sqrt{1+\mu}\sqrt{n+1}|n+1\rangle\, .$$
This is perfectly consistent with the trivial background treated in section 2, since in that
case, to leading order in $N$, we have ${c\over N}=1$. We can consider the action of a derivative on the operator ${\cal O}_n$.
It is straight forward to verify that (the rule for the derivative of a Schur polynomial was conjectured in \cite{related} and proved
in \cite{de Mello Koch:2007uu})
$$\left[ {\Tr {d\over dZ}\over\sqrt{N}}\, ,\,{1\over (1+\mu)^{n/2}\sqrt{n!}}\left({\chi_{\yng(1)}(Z)\over\sqrt{N}}\right)^n{\chi_B(Z)\over\sqrt{f_B}}\right]$$
\begin{eqnarray}
&=&\sqrt{n}\sqrt{1+\mu}\left[{1\over (1+\mu)^{(n-1)/2}\sqrt{(n-1)!}}\left({\chi_{\yng(1)}(Z)\over\sqrt{N}}\right)^{n-1}{\chi_B(Z)\over\sqrt{f_B}}\right]
+\cdots,\nonumber\\
&=&\sqrt{n}\sqrt{1+\mu}{\cal O}_{n-1}+\cdots\quad\label{puzzle}
\end{eqnarray}
where $\cdots$ represents an extra term that does not yet have an obvious interpretation\footnote{This
extra term arises from the action of the derivative on $\chi_B(Z)$. The term that we have written explicitely
removes some of the boxes that $(a^\dagger )^n$ added, which is the expected action of an annihilation
operator.}. Ignoring this term, we can identify, 
$$ {\Tr {d\over dZ}\over\sqrt{N}}\leftrightarrow a,$$
where the graviton annihilation operator acts as
$$ a|n\rangle =\sqrt{c\over N}\sqrt{n}|n-1\rangle =\sqrt{1+\mu} \sqrt{n}|n-1\rangle \, .$$
Clearly $\big[ a,a^\dagger\big]=1+\mu$. However, this can't be correct: the identification
$$\big[ {1\over\sqrt{N}}\Tr \left({d\over dZ}\right),{\Tr (Z)\over\sqrt{N}} \big]\leftrightarrow\big[ a,a^\dagger \big]=1+\mu ,$$
implies that
$$\big[ {1\over\sqrt{N}}\Tr \left({d\over dZ}\right),{\Tr (Z)\over\sqrt{N}} \big]=1+\mu,$$
which is in stark disagreement with (\ref{comm}). To resolve this discrepancy, note that in contrast to the trivial background considered in the previous 
section, we can now act on the operator dual to the vacuum with the derivative operator to obtain a non-zero result\footnote{In the previous section 
the operator dual to the vacuum was the identity.}. Indeed, we find
$$ \left({\Tr ({d\over dZ})\over\sqrt{N}}\right)^n {\chi_B(Z)\over\sqrt{f_B}}
=\sum_{R}d_R\, \prod_{i}c_i\, {\chi_{- R} (Z)\over \sqrt{f_B N^n}} \, . $$
In the last line, $\prod_{i}c_i$ is the product of the weights of the boxes removed by the derivatives. This formula is again exact.
The two point correlator of this operator, to the leading order in the large $N$ limit, is 
$$ \left\langle \left[\left({\Tr ({d\over dZ})\over\sqrt{N}}\right)^n {\chi_B(Z)\over\sqrt{f_B}}\right]^\dagger\left({\Tr ({d\over dZ})\over\sqrt{N}}\right)^n 
{\chi_B(Z)\over\sqrt{f_B}}\right\rangle\, =\, \mu^n\, n! . $$
A natural interpretation is that the derivative operator (which lowers the total ${\cal R}$ charge of the state) creates gravitons moving in the opposite
direction, which we will call {\it countergravitons}\footnote{We
call a state with positive (negative) ${\cal R}$-charge $Q=\Tr Z {d\over dZ}$ after subtracting off the ${\cal R}$-charge of the vacuum a graviton
(countergraviton). The term {\sl antigraviton} would be incorrect - antigravitons are created by $Z^\dagger$; including any antigravitons in the state
would take us out of the ${1\over 2}$ BPS sector.}. An operator dual to a normalized $n$-countergraviton state, to leading order in $N$, is now given by
$$ {\cal O}_{-n} ={1\over \mu^{n/2}\sqrt{n!}}\left({\Tr \left({d\over dZ}\right)\over\sqrt{N}}\right)^n{\chi_B(Z)\over\sqrt{f_B}} .$$
Denote the countergraviton creation and annihilation operators by $b^\dagger$ and $b$. In this case
$$ \big[{\Tr ({d\over dZ})\over\sqrt{N}},{\cal O}_{-n}\big] =\sqrt{c\over N}\sqrt{n+1}{\cal O}_{-n-1}=\sqrt{\mu}\sqrt{n+1}{\cal O}_{-n-1} \, ,$$
(where $c$ is the weight of the box removed from the background Young diagram $B$ after acting with ${d\over dZ}$) becomes
$$ b^\dagger |-n\rangle =\sqrt{\mu}\sqrt{n+1}|-n-1\rangle \, ,$$
in the dual quantum gravity. Note that we are using the field theory ${\cal R}$-charge (string theory angular momentum) to label states. 
Arguing exactly as above, we have $\big[b,b^\dagger\big]=\mu$.
The resolution to our paradox is now in sight: the derivative operator creates countergravitons and destroys gravitons; multiplication
by $\Tr (Z)$ creates gravitons and destroys countergravitons. It is now easy to argue for the 
identifications\footnote{These identifications also provide an interpretation for the extra term in 
equation (\ref{puzzle}): the extra term corresponds to a state with $n$ gravitons and 1 counter graviton.}
$$ {\Tr (Z)\over\sqrt{N}}\leftrightarrow a^\dagger + b \, ,\qquad {\Tr ({d\over dZ})\over\sqrt{N}}\leftrightarrow a + b^\dagger \, .$$
These identifications are perfectly consistent with (\ref{comm}). Indeed,
$$\big[ {1\over\sqrt{N}}\Tr \left({d\over dZ}\right),{\Tr (Z)\over\sqrt{N}} \big]\leftrightarrow\big[ a+b^\dagger,a^\dagger +b\big]=(1+\mu )-(\mu)=1\, .$$
Thus, we see that $\Tr (Z)$ and $\Tr\left({d\over dZ}\right)$ no longer have a local action on the Young diagram $B$.
We will define new operators that are local: split
\begin{equation} 
\Tr (Z)=\Tr (Z)_a + \Tr (Z)_b .
\label{split}
\end{equation}
The definition of $\Tr (Z)_a$ is that it can add boxes to $B$, but only in the upper right hand corner of the Young diagram, corresponding
to the action of $a^\dagger$; $\Tr (Z)_b$ can add boxes to $B$, but only in the lower right hand corner of the Young diagram, corresponding
to the action of $b$. Similarly, split
$$ \Tr \left({d\over dZ}\right)= \Tr \left({d\over dZ}\right)_a + \Tr \left({d\over dZ}\right)_b .$$
$\Tr \left({d\over dZ}\right)_a$ removes boxes from the upper right hand corner corresponding to the action of $a$ and
$\Tr \left({d\over dZ}\right)_b$ removes boxes from the lower right hand corner corresponding to the action of $b^\dagger$.
Diagrammatically we have
$$ \Tr (Z)_a\chi_{\tiny \yng(4,4,4,4,3)}=\chi_{\tiny \yng(5,4,4,4,3)},\qquad \Tr (Z)_b\chi_{\tiny \yng(4,4,4,4,3)}=\chi_{\tiny \yng(4,4,4,4,4)},$$
$$ \Tr \left({d\over dZ}\right)_a\chi_{\tiny \yng(5,4,4,4,4)}=(N+4)\chi_{\tiny \yng(4,4,4,4,4)},\qquad 
\Tr \left({d\over dZ}\right)_b\chi_{\tiny \yng(5,4,4,4,4)}=(N-1)\chi_{\tiny \yng(5,4,4,4,3)} \, .$$
We are computing correlation functions using the product rule and known two point functions for Schur polynomials. We have just defined
the product of the new local operators with the usual Schur polynomials. This product gives a linear combination of Schur polynomials
whose two point function is known. Thus we can compute any correlators using the new local operators which shows that these local
operators are indeed well enough defined for our purposes. We will give a more constructive definition of these local operators below.

We can again use the operators we have defined to build a graviton coherent state and check that we reproduce the expected graviton dynamics.
The operator dual to a coherent graviton state is
$$ {\cal O}_z = {\cal N}\sum_{n=0}^\infty {z^n\, {\cal O}_n\over (1+\mu)^{n\over 2}\sqrt{n!}}\, \qquad {\rm where}$$  
$$ {\cal O}_n ={1\over (1+\mu)^{n/2}\sqrt{n!}}\left({\Tr(Z)_a\over\sqrt{N}}\right)^n{\chi_B(Z)\over\sqrt{f_B}} .$$
Requiring a unit two point function determines the normalization
$$ \left\langle {\cal O}_z^\dagger {\cal O}_z\right\rangle =|{\cal N}|^2 e^{|z|^2\over 1+\mu}=1,$$
$$ {\cal O}_z = e^{-|z|^2\over 2(1+\mu)}\sum_{n=0}^\infty {z^n\, {\cal O}_n\over (1+\mu)^{n\over 2}\sqrt{n!}}\, .$$
This operator satisfies
$$ \left[ {\Tr\left({d\over dZ}\right)_a\over \sqrt{N}},{\cal O}_z\right]=z{\cal O}_z\, .$$
Denote the coherent state that this operator maps to by $|z\rangle$. The Hamiltonian is
\begin{equation}
H={a^\dagger a\over 1+\mu}-{b^\dagger b\over \mu}.
\label{ham}
\end{equation}
This Hamiltonian measures the energy with respect to the background $B$. It is for this reason that
the second term above is negative: the countergravitons are missing boxes, similar to holes in the Dirac sea.
Since the sea is finite the Hamiltonian is bounded below.

To compare results to the dual gravity, we need to translate the background $B$ into an LLM geometry, using the 
results of \cite{Lin:2004nb} which we will now review. This is most easily done by first translating the Schur 
polynomial into a free fermion state and then translating this fermion state into a boundary condition for the
LLM geometry. The Schur polynomials map into an $N$ fermion energy eigenstate\cite{Corley:2001zk,Berenstein:2004kk}, 
which is labeled by a set of $N$ integers telling us which levels are occupied. 
Denote the number of boxes in row $i$ of the Young diagram by $r_i$; the row closest to the bottom of the page has $i=1$ and the
row closest to the top of the page has $i=N$. The energies of the occupied levels are given by $r_i+i-1$. These occupied states
map into rings in the free fermion phase space\footnote{Recall that our fermions are in an external $x^2$ potential so that fixed
energies $E=p^2+x^2$ are rings in the $x,p$ plane.}. To get a correspondence with ${1\over 2}$ BPS geometries, 
identify the free fermion phase space with the LLM plane\footnote{We call the two dimensional plane on which the LLM boundary condition is 
specified the LLM plane. This plane is at $y=0$ and has coordinates $x^1 , x^2$.} and color occupied regions in black 
and unoccupied regions in white. This black and white pattern is the boundary 
condition for the dual geometry. As an example, we have shown the boundary condition dual to $B$ in figure 5 below.

\myfig{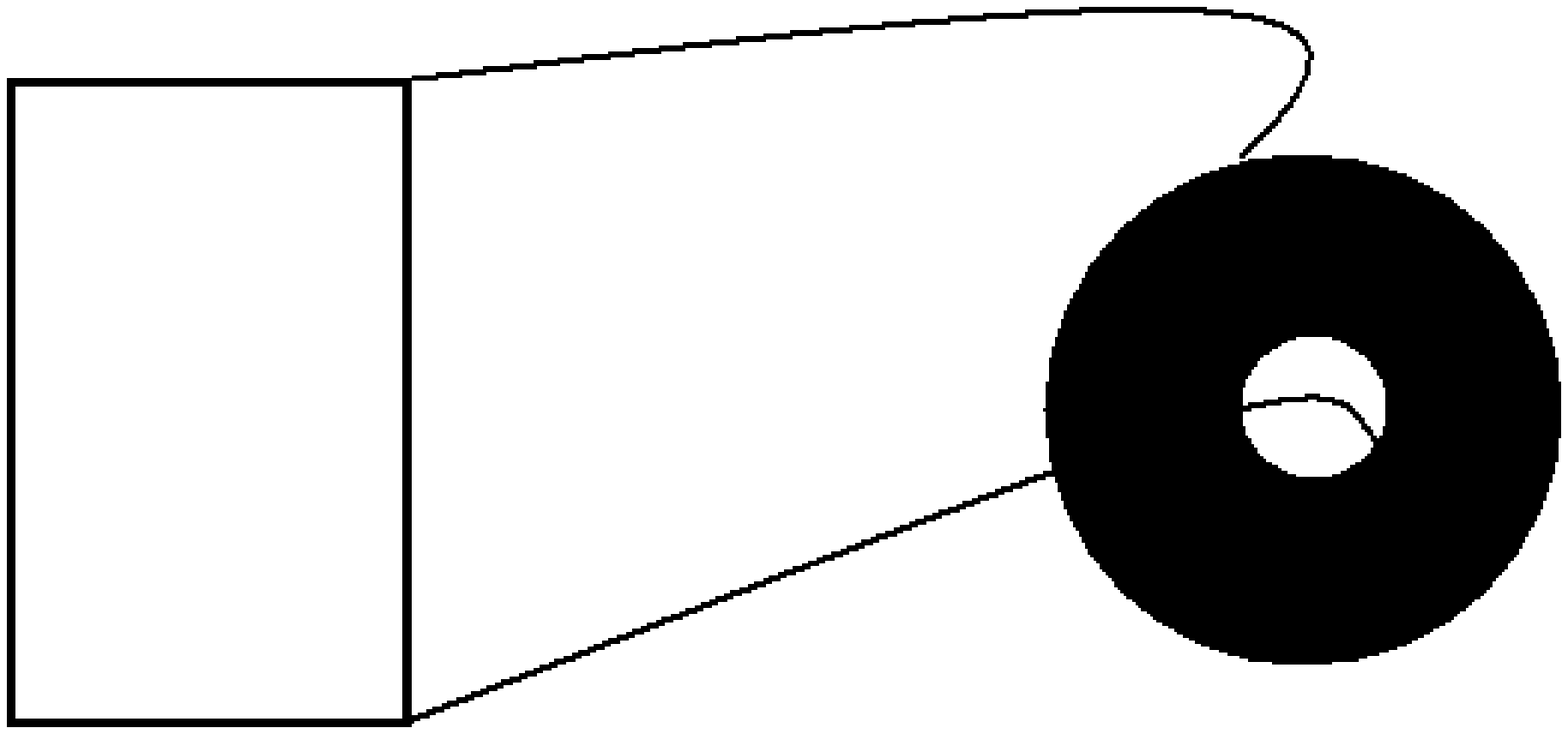}{10.0}{The Young diagram $B$ is shown on the left. The corresponding boundary condition for the LLM geometry is shown on the right.}

By acting with the graviton creation and annihilation operators, we are producing small ``ripples'' on the inner and outer edges of the annulus. 
These ripples are localized in the radial direction of the LLM plane in spacetime and they are localized on the Young diagram. We see a direct
connection between locality in spacetime and locality on the Young diagram. This is a feature of all of the examples we have considered.
We have seen that the oscillators in the new background are scaled by a factor of ${1\over N}$ times the weight of the box removed/added by the
oscillator. Does this factor have an interpretation in the dual geometry? The area of the black regions on the LLM plane are proportional to
the number of states in the region. We will use a normalization for which the area divided by $\pi/N$ is exactly equal to the number of
states in the black region. Thus, the total area of the black regions is $\pi $. The AdS$_5\times $S$^5$ background itself corresponds to a black disc
of radius $R=1$, in a sea of white. The Young diagram $B$ has a total of $N$ rows and $M$ columns, so that in the free fermion description, there are $M$ 
unoccupied states followed by $N$ occupied states giving a phase space that looks like an annulus. The hole in the center of the annulus has
a total of $M$ states and hence a radius of $\sqrt{M\over N}=\sqrt{\mu}$. Note that $\mu$ is precisely equal to the 
weight of the box in the lower right hand corner divided by $N$. The area of the black annulus and its white center is proportional to $M+N$
(since together they correspond to $M+N$ states). The outer radius of the annulus is thus $\sqrt{N+M\over N}=\sqrt{1+\mu}$. 
Note that $1+\mu$ is precisely equal to the weight of the box in the upper right hand corner divided by $N$. Thus, the factor by which 
the oscillators are rescaled is equal to the value of radius in the LLM plane at which the dual graviton is created.

What is the energy of a state of gravitons created at a given radius on the LLM plane? Given the interpretation 
in terms of a free fermion phase space, this energy is just the number of gravitons times the energy of a single 
graviton and the energy of a single graviton is given by the radius squared times $N$. This tells us that gravitons are 
small ripples on the outer edge of the annulus and countergravitons are ripples on the inner edge of the annulus.

We will now focus on excitations of the outer edge of the annulus, i.e. those created by $a^\dagger$. 
The energy measured by (\ref{ham}) is measured with respect to the background $B$.
In the LLM picture, gravitons exciting the outer edge of the annulus have an energy $N(1+\mu)$. This is the
energy of the graviton measured with respect to the true vacuum of the theory. There is a natural way
to modify (\ref{ham}) to get agreement with the LLM result: we can introduce a rescaled time coordinate 
$\tau=N(1+\mu)t$. In this case, the Hamiltonian becomes
$$ H= N a^\dagger a .$$
To see that this rescaling does the job, note that the expectation value of the energy of a coherent state
of $a^\dagger$ quanta is
$$ \langle z|H|z\rangle=\langle z|a^\dagger a|z\rangle=\bar{z}z= N r^2 .$$ 
The number operator acts as
$$\big[ N,{\cal O}_z\big]= e^{-|z|^2\over 2(1+\mu)}\sum_{n=0}^\infty {n\, z^n\, {\cal O}_n\over (1+\mu)^{n\over 2}\sqrt{n!}}\, ,$$
so that the expected number of gravitons, is given by 
$$ \langle z|N|z\rangle = \langle\big[ N,{\cal O}_z\big]{\cal O}_z^\dagger\rangle = {r^2\over 1+\mu}.$$
Comparing the expectation value of the energy to the expectation value of the number operator
we see that each graviton has an energy of $N(1+\mu)$. The Lagrangian governing the low energy 
excitations of this coherent state is
$$ L=\langle z|i{d\over dt}|z\rangle -\langle z|H|z\rangle = N\left({d\phi\over d\tau} r^2 - r^2\right) .$$
This is indeed the Lagrangian for the dynamics of
gravitons in the LLM geometry (we have reproduced (3.41) of \cite{Mandal:2005wv}). Notice also that the angular momentum
($={\cal R}$ charge) computed using the rescaled time $\tau$
$$ {\partial L\over \partial{d\phi\over d\tau}}=N r^2 $$
is equal to the energy, as expected for a BPS state. 
A similar argument for a coherent state of countergravitons gives\footnote{In this case, we have introduced a
different rescaled time coordinate  $\tau=\mu t$.}
$$ \langle z|H|z\rangle =\langle z|b^\dagger b |z\rangle =  -Nr^2, \qquad \langle z|N|z\rangle = {r^2\over\mu} .$$
This suggests that each countergraviton has an energy of $N \mu$. This matches the energy of a graviton
exciting the inner edge of the annulus. 

It is not surprising that we need to work with the energy measured with respect to the true vacuum, to obtain agreement with
the dual gravity picture. The giant graviton probe of \cite{Mandal:2005wv} is identified with a highly excited fermion; the
energy measured with respect to the true vacuum is precisely the energy of this fermion. In addition, the analysis of
\cite{Balasubramanian:2005mg} focuses on very heavy half-BPS states. Almost all such states correspond to Young diagrams 
whose shapes are small fluctuations around a fixed limiting curve. This characterization of typical states is done using
statistical ensembles which assign the energy measured with respect to the true vacuum, to each state.
In both cases above the rescaled time coordinate is given by $\tau= c t$ with $c$ the weight of
the boxes that are added or removed. Thus, we are lead rather naturally to rescaling time 
differently for the various sectors of the field theory. The LLM geometries have a time coordinate that is warped, so that 
there are various ``local'' times as well as the global time coordinate. Our work might provide a natural way to understand 
these time rescalings in the field theory\footnote{We would like to thank the anonymous referee who suggested
this interpretation.}. We hope to return to this issue.

Using Schur technology it is simple to verify that
\begin{equation}
\left\langle {\chi_B(Z)^\dagger\over\sqrt{f_B}} \left({\Tr (Z^{p\, \dagger} )\over\sqrt{pN^p}}\right)^n\left({\Tr (Z^p)\over\sqrt{pN^p}}\right)^m 
{\chi_B(Z)\over\sqrt{f_B}}\right\rangle = n! \big[ (1+\mu )^p\big]^n\delta_{m,n}\, .
\label{multimom}
\end{equation}
Thus, we can again identify operators which create gravitons which each carry $p$ units of angular momentum. Does the graviton operator $\Tr (Z^p)$
have a local decomposition? We will now define and give evidence for the {\sl matrix replacement} $Z\to Z_a+Z_b$. Using the results of 
\cite{Balasubramanian:2004nb,de Mello Koch:2007uu,de Mello Koch:2007uv}, we know that we 
can associate matrices in the Schur polynomial with boxes in the Young diagram.
Using this insight, at the level of the Schur polynomial, the replacement amounts to allowing each box to have an identity `a' or `b' and summing
over all possible identities. Thus, for example
$$ \Tr (Z^2)=\chi_{\tiny \young(aa)}(Z)+\chi_{\tiny \young(ab)}(Z)+\chi_{\tiny \young(ba)}(Z)+\chi_{\tiny \young(bb)}(Z)
-\chi_{\tiny \young(a,a)}(Z)-\chi_{\tiny \young(a,b)}(Z)-\chi_{\tiny \young(b,a)}(Z)-\chi_{\tiny \young(b,b)}(Z). $$
When we now take a product of Young diagrams, we use the usual Littlewood-Richardson rule except that we are only allowed to add `a' boxes
to the upper right hand region of the background Young diagram and `b' boxes to the lower 
right hand region of the background Young diagram. This refinement of
the Littlewood-Richardson rule defines what we mean by the decomposition $Z\to Z_a+Z_b$. We have checked that this recipe reproduces
(\ref{multimom}). Thus, for example, $\Tr (Z_a^p)$ creates a localized graviton, localized at the outer edge of the annulus, carrying $p$
units of angular momentum.

Note that, to leading order in $N$, when $B$ has $O(N^2)$ boxes and $\sum_i p_i$ is $O(1)$, we have (the $p_i$ are all distinct)
$$
\left\langle {\chi_B(Z)^\dagger\over\sqrt{f_B}} \prod_i \left({\Tr (Z^{p_i\, \dagger} )\over\sqrt{p_i N^{p_i}}}\right)^n
\left({\Tr (Z^{p_i})\over\sqrt{p_iN^{p_i}}}\right)^m 
{\chi_B(Z)\over\sqrt{f_B}}\right\rangle \equiv \left\langle  \prod_i \left({\Tr (Z^{p_i\, \dagger} )\over\sqrt{p_i N^{p_i}}}\right)^n
\left({\Tr (Z^{p_i})\over\sqrt{p_iN^{p_i}}}\right)^m \right\rangle_B $$
$$=  \prod_i \left\langle  \left({\Tr (Z^{p_i\, \dagger} )\over\sqrt{p_i N^{p_i}}}\right)^n
\left({\Tr (Z^{p_i})\over\sqrt{p_iN^{p_i}}}\right)^m \right\rangle_B .
$$
The fact that the expecation value of the product is the product of expectation values, where we define expectations by
$$  \left\langle  O \right\rangle_B \equiv  \left\langle {\chi_B(Z)^\dagger\over\sqrt{f_B}} O {\chi_B(Z)\over\sqrt{f_B}} \right\rangle $$
is in fact a consequence of the fact that $\chi_B(Z)$ creates a new background. Indeed, whenever
the expectation value of the product is the product of expectation values, we are dealing with a classical limit.

We have considered excitations that correspond to small ripples at the edges of the annulus. These excitations
correspond to point like gravitons and their ${\cal R}$ charge (with respect to the background) is $O(1)$. We can also consider 
excitations with a large ${\cal R}$-charge, of $O(N)$. These excitations correspond to giant gravitons.
\myfig{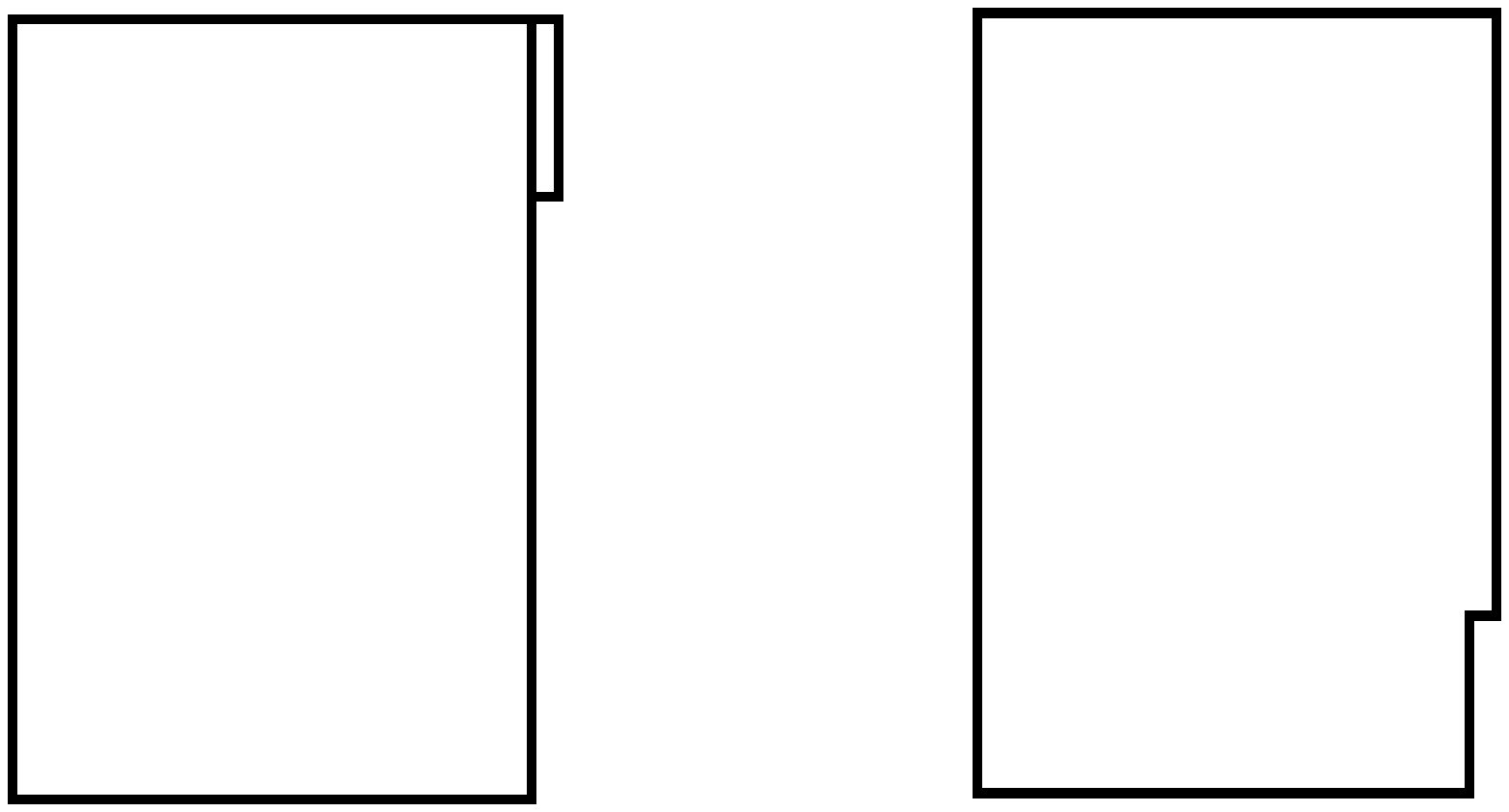}{8.0}{The Young diagrams corresponding to giant gravitons.}

To create a giant graviton, we need to build up a single column till it has $O(N)$ boxes stacked in the column; to create a counter giant graviton
we need to remove boxes from a single column till we have a scar of $O(N)$ boxes eroded from the Young diagram. These two possibilities are shown in figure
6. Denote the number of boxes added/eroded to produce the giant by $p$. Set $\mu_p={p\over N}$. When we build our coherent state below, we will assume that
all giant graviton states contributing to the coherent state have $p+O(1)$ boxes. Thus, the weight of the last box added to produce the giant divided
by $N$ is $1+\mu-\mu_p$ and the weight of the last box eroded to produce the countergiant divided
by $N$ is $\mu+\mu_p$. Using the methods we introduced above, we can write
$$ {\Tr (Z)\over\sqrt{N}}\leftrightarrow a^\dagger + b +A^\dagger +B \, ,\qquad 
   {\Tr ({d\over dZ})\over\sqrt{N}}\leftrightarrow a + b^\dagger + A + B^\dagger \, .$$
$a ,b ,a^\dagger ,b^\dagger$ are the operators introduced above which satisfy
$$\big[a,a^\dagger \big]=1+\mu,\qquad \big[b,b^\dagger\big]=\mu .$$
The operator $A^\dagger$ adds one unit of energy to the giant; $B$ removes one unit of energy from the counter giant. $A$ and $B$ are not normal
bosonic operators - they are Cuntz oscillators, satisfying
$$ AA^\dagger = {c\over N}=1+\mu-\mu_p,\qquad A^\dagger A = (1+\mu-\mu_p)(1-|0\rangle\langle 0|),$$
$$ BB^\dagger = {c\over N}=\mu+\mu_p,\qquad B^\dagger B = (\mu+\mu_p)(1-|0\rangle\langle 0|).$$
In these formulas, $c$ is again the weight of the box which is added/removed by the oscillator. An operator dual to
a normalized giant graviton state with energy $n$ is given by
$$ {\cal O}_p=\sqrt{N^p (N+M-p)!\over (N+M)!}\left( {\Tr (Z)_A\over\sqrt{N}}\right)^p {\chi_B(Z)\over\sqrt{f_B}}; $$
an operator dual to a giant graviton coherent state is given by
$$ {\cal O}_z={\cal N}\sum_{n=0}^\infty {z^n\over (1+\mu)_{n}}\left( {\Tr (Z)_A\over\sqrt{N}}\right)^n {\chi_B(Z)\over\sqrt{f_B}}\qquad
(1+\mu)_n=\prod_{i=1}^n (1+\mu+\mu_i),$$
$$ {\cal N}=\left[
\sum_{n=0}^\infty \left({|z|^n\over (1+\mu)_n}\right)^2 {(N+M)!\over (N+M-n)! \, N^n}
\right]^{1\over 2}\, .$$
We take $z$ small enough that the expected number of gravitons in this state $p=O(N)$ and $\mu_p\ll 1$.
It is straight forward to verify that
$$ \left[ {\Tr\left( {d\over dZ}\right)\over\sqrt{N}},{\cal O}_p\right]=\sqrt{1+\mu-\mu_p}{\cal O}_{p-1},\qquad
\left[ {\Tr\left( {d\over dZ}\right)\over\sqrt{N}},{\cal O}_z\right]=z{\cal O}_z .$$
The Lagrangian governing the low energy excitations of this giant graviton coherent state is\footnote{$t=-2N\tau{d\over dr^2}\log {\cal N}$; since the
equations of motion set $r$ to a constant we have treated $r$ as a constant in performing this rescaling.}
$$ L=\langle z|i{d\over dt}|z\rangle - \langle z|H|z\rangle =N\left(r^2{d\phi\over d\tau}-r^2\right) .$$
which is the correct Lagrangian for D3 branes in an LLM geometry\cite{Mandal:2005wv}.

Finally, consider probing the background $B$ with a string. The same computation, using different methods,
has been given in \cite{Chen:2007gh}. We probe with a closed string of the form
$$\Tr (YZ^{n_1}_a YZ^{n_2}_a Y\cdots YZ^{n_L}_a), $$
corresponding to a string localized at the outer edge of the annulus, or
$$\Tr (Y\left({d\over dZ}\right)^{n_1}_b Y\left({d\over dZ}\right)^{n_2}_b Y\cdots Y\left({d\over dZ}\right)^{n_L}_b), $$
corresponding to a string localized on the inner edge of the annulus. These loops are not ${1\over 2}$-BPS and hence their
one loop anomalous dimensions are non-zero. One can define a Cuntz oscillator Hamiltonian whose spectrum gives this one-loop anomalous dimension.
The computation of the relevant correlators involving these loops is outlined in appendix B. In the AdS$_5\times$S$^5$ 
background, these correlators can be computed by associating a lattice with the $Y$ Higgs fields and Cuntz oscillators with the $Z_a$s and 
$\left({d\over dZ}\right)_b$s. In the background $B$ this is also the case, except that the Cuntz oscillators are rescaled by the 
(now familiar) square root of the weight divided by $N$. For $Z_a$
we use
$$ \alpha \alpha^\dagger =1+\mu,\qquad  \alpha^\dagger\alpha =(1+\mu)(1-|0\rangle\langle 0|),$$
and for $\left({d\over dZ}\right)_b$ we use
$$ \beta \beta^\dagger =\mu,\qquad  \beta^\dagger\beta =\mu(1-|0\rangle\langle 0|).$$
The Cuntz oscillator Hamiltonian which gives the one loop anomalous dimension, at large $N$, is (see appendix B)
$$ H=\lambda\sum_{l=1}^L (\alpha_l-\alpha_{l+1})^\dagger (\alpha_l-\alpha_{l+1})$$
for strings localized at the outer edge of the annulus. In the large $N$ limit, where our Cuntz operator representation is valid, the operator
$$ {\cal O}_z =\sum_{n=0}^\infty {z^n\over (1+\mu)^{n}}Z_a^n = \sum_{n=0}^\infty {z^n\over (1+\mu)^{n}}(\alpha^\dagger)^n ,$$
is dual to a coherent state. Indeed, it satisfies $\big[ \alpha , {\cal O}_z\big] = z {\cal O}_z$. 
We can put each site of the loop into such a coherent state
with a different coherent state parameter; denote the resulting state by $|z_1,z_2,\cdots,z_L\rangle .$
The semiclassical sigma model action governing the low energy dynamics of this lattice model is given by\footnote{$H$ in this action does not include
the leading contribution to the dimension of the operator; it includes only the $g_{YM}^2$ correction. For the loop we consider the full contribution comes from
the $F$-term. See \cite{Vazquez:2006id,Chen:2007gh} and references therein.}
$$ S=\int dt\left(i\langle z_1,z_2,\cdots,z_L|{d\over dt}|z_1,z_2,\cdots,z_L\rangle -
\langle z_1,z_2,\cdots,z_L|H|z_1,z_2,\cdots,z_L\rangle \right).$$
In the large $L$ limit the above sums become integrals leading to
$$ S=L\int dt\int_0^1 d\sigma \left({(1+\mu)r^2\dot{\phi}\over 1+\mu-r^2}-{\lambda\over L^2}\partial_\sigma z
\partial_\sigma\bar{z}\right).$$
Comparing this to the fast string limit of the Polyakov action for a string moving in an LLM geometry, we read off
(recall that the outer radius of the annulus $R_o=\sqrt{1+\mu}$)
$$ V_\phi (x_1,x_2,y=0)= {1+\mu\over 1+\mu-r^2}={R_o^2\over R_o^2-r^2},$$
which is the correct result for $V_\phi (x_1,x_2,y=0)$ as felt by a string moving at the outer edge of the annulus. Probing the
inner edge gives (recall that the inner radius of the annulus $R_i=\sqrt{\mu}$)
$$ V_\phi (x_1,x_2,y=0)= -{\mu\over \mu-r^2}=-{R_i^2\over R_i^2-r^2}.$$
The negative sign in this formula comes from the fact that when we probe the inner edge we do so with counter excitations.
Summing these two gives the complete LLM answer for $V_\phi (x_1,x_2,y=0)$.

\subsection{An Arbitrary Number of Rings}

We can generalize the results of the previous section to consider the Young diagram shown in figure 7. The geometry now corresponds to a 
sequence of concentric rings. 
\myfig{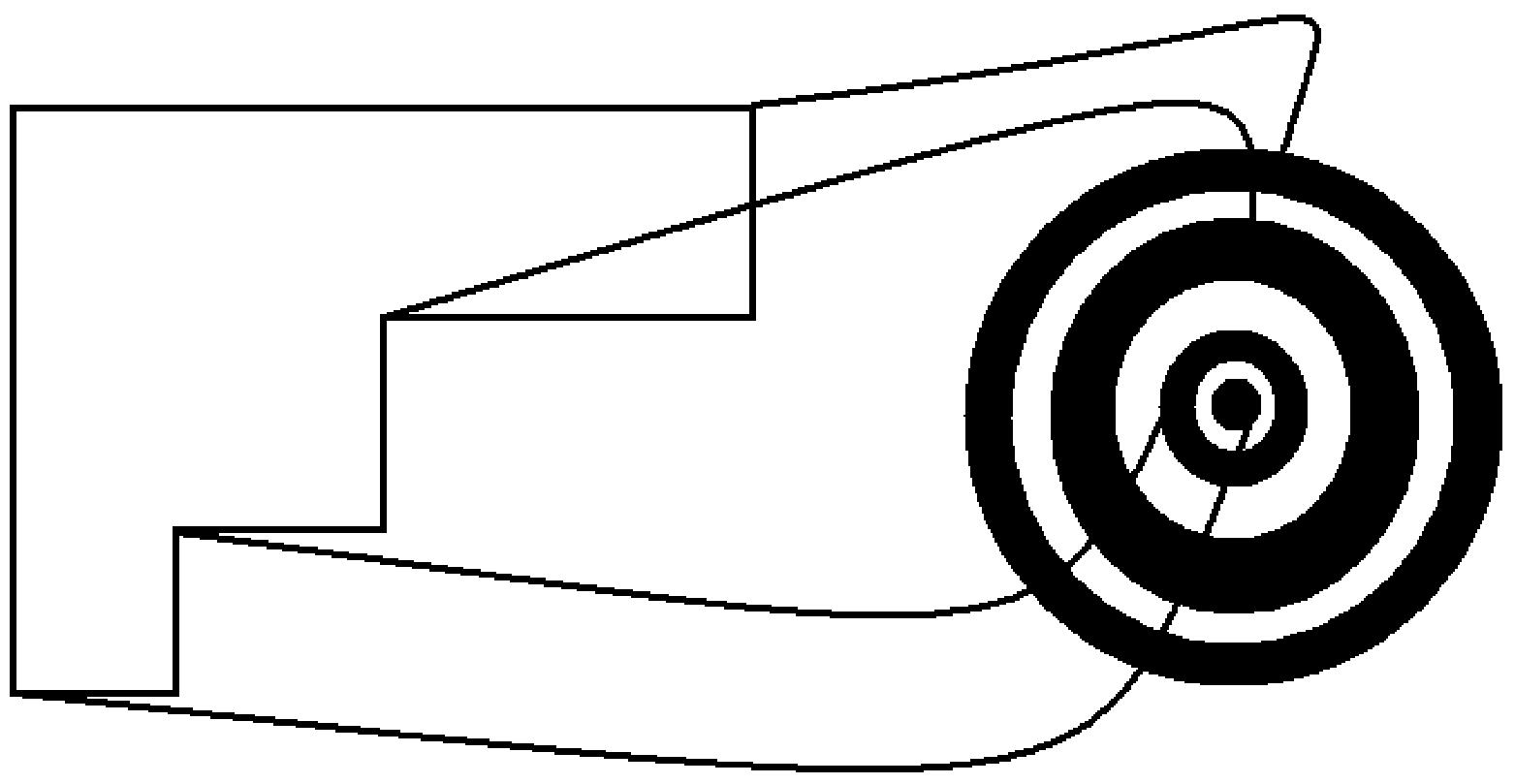}{12.0}{This Young diagram is a generalization of the background considered in the previous section. The outer
edge of each ring corresponds to an inward pointing corner.}

A corner of the Young diagram that we can erode by removing blocks is called an outward pointing corner; a corner of the Young diagram
that can be used to grow the diagram by adding blocks is called an inward pointing corner. Fields localized at an outward (inward)
pointing corner carry a subscript $b$ ($a$). The corners identified in figure 7 all correspond 
to inward pointing corners. ${\Tr (Z)\over\sqrt{N}}$ can be decomposed as
$${\Tr (Z)\over\sqrt{N}}=\sum_{i}{\Tr (Z_{a_i})\over\sqrt{N}}+\sum_{j}{\Tr (Z_{b_j})\over\sqrt{N}},$$
where $i$ is summed over all inward pointing corners and $j$ is summed over all outward pointing corners. Each of the local fields on the 
right hand side can only add boxes at their corner. ${\Tr (Z)\over\sqrt{N}}$ corresponds, in the dual quantum gravity,
to a sum of oscillator $\sum_i a^\dagger_i +\sum_j b_j$. Similarly,
$${1\over\sqrt{N}}\Tr\left({d\over dZ}\right)=\sum_i {1\over\sqrt{N}}\Tr\left({d\over dZ_{a_i}}\right)+
\sum_j {1\over\sqrt{N}}\Tr\left({d\over dZ_{b_j}}\right)$$
corresponds to the sum of oscillator $\sum_i b^\dagger_i +\sum_j a_j$ where again $i$ is summed over all outward pointing corners and $j$ 
is summed over all inward pointing corners. It is a simple matter to check that
$$\big[ a_i,a_i^\dagger\big]={c_{a_i}\over N},$$
where $c_{a_i}$ is the weight of the boxes at the $i$th inward pointing corner and
$$\big[ b_j,b_j^\dagger\big]={c_{b_j}\over N},$$
where $c_{b_j}$ is the weight of the boxes at the $j$th outward pointing corner. Finally, arbitrary mixed correlators, for example 
of the form
$$\left\langle {\chi_B(Z)\over\sqrt{f_B}}\Tr \left((Z_{a_i})^n(Z^\dagger_{a_i})^n\right){\chi_B(Z)^\dagger\over\sqrt{f_B}}\right\rangle
\equiv \left\langle \Tr \left((Z_{a_i})^n(Z^\dagger_{a_i})^n\right)\right\rangle_B $$
$$\left\langle \left[\left({d\over dZ_{b_i}}\right)^n\right]_{ij}{\chi_B(Z)\over\sqrt{f_B}}
\left[\left({d\over dZ^\dagger_{b_i}}\right)^n\right]_{ji}{\chi_B(Z)^\dagger\over\sqrt{f_B}}\right\rangle
\equiv \left\langle \Tr \left(\left({d\over dZ_{b_i}}\right)^n \left({d\over dZ^\dagger_{b_i}}\right)^n\right)\right\rangle_B $$
are easily computed using the modified ribbon graphs
$$ \left\langle (Z_{a_i})_{ij}(Z^\dagger_{a_i})_{kl}\right\rangle_B = {c_{a_i}\over N}\delta_{il}\delta_{jk},$$
$$ \left\langle \left({d\over dZ_{b_j}}\right)_{ij}\left({d\over dZ^\dagger_{b_j}}\right)_{kl}\right\rangle_B = {c_{b_j}\over N}\delta_{il}\delta_{jk}.$$

\subsection{Eigenvalues}

The split of both $\Tr(Z)$ and $\Tr\left({d\over dZ}\right)$ into local operators has been defined by stating what action these operators
have on the background Young diagram. Is there a more direct definition which would allow us to construct these 
local (in the bulk) operators? The
${1\over 2}$ BPS sector of the theory on $R\times S^3$ is governed by the singlet sector dynamics of the action
$$ S=\int dt\, \Tr\left({1\over 2}{dZ\over dt} {dZ^\dagger\over dt}-{1\over 2}ZZ^\dagger\right).$$
The eigenvalue dynamics of this model is the dynamics of $N$ nonrelativistic 
fermions\cite{Corley:2001zk,Berenstein:2004kk,Balasubramanian:2005mg,Takayama:2005yq}. 
The eigenvalues of $Z$ are related to the positions of the fermions.
Now, recall a few facts from the usual harmonic oscillator: the position $X\sim a^\dagger+a$ has a vanishing expectation value for any energy
eigenstate; for $X^2$ however we find
$$ \langle n|(a+a^\dagger)^2|n\rangle = \langle n|(aa^\dagger + a^\dagger a)|n\rangle \sim n,$$
so that the expectation value of $X^2$ is large if $n$ is large.

In the large $N$ limit, we expect that the eigenvalues of $Z$ can be described by a classical configuration. Armed with our harmonic oscillator
insight, we expect that the eigenvalues will be large if the corresponding fermion is highly excited. A highly excited fermion corresponds to a
long row in the Young diagram. Thus, the Young diagram of the background $B$ is a ``picture'' of the eigenvalues of the classical large $N$
configuration for $Z$ in the sense that there is an eigenvalue for each row in the Young diagram and the size of the eigenvalue is related to
the length of the row. For example, the background corresponding to the Young diagram pictured in figure 7 will have eigenvalues
clumped around four values. $Z$ can thus be split into four submatrices which each have eigenvalues of roughly the same size. We suggest that this split
is precisely the split of $Z$ into local operators.

It is easy to gather numerical support for this proposal. Consider a matrix
$$ Z=\left[\matrix{\lambda_1 &0\cr 0 &\lambda_2}\right]. $$
The product
$$\chi_{\tiny\yng(3)}(Z)\chi_{\tiny\yng(1)}(Z)=\chi_{\tiny\yng(3,1)}(Z)+\chi_{\tiny\yng(4)}(Z)$$
follows as a simple application of the Littlewood-Richardson rule. Our local operators will satisfy
$$\chi_{\tiny\yng(3)}(Z)\chi_{\tiny\young(a)}(Z)=\chi_{\tiny\yng(4)}(Z),\qquad 
  \chi_{\tiny\yng(3)}(Z)\chi_{\tiny\young(b)}(Z)=\chi_{\tiny\yng(3,1)}(Z).$$
Consider the case that $\lambda_1\gg\lambda_2$. Then our proposal says $Z=Z_a+Z_b$ with
$$ Z_a=\left[\matrix{\lambda_1 &0\cr 0 &0}\right],\qquad Z_b=\left[\matrix{0 &0\cr 0 &\lambda_2}\right]. $$
We thus have the predictions
$$\chi_{\tiny\yng(3)}(Z)\lambda_1 =\chi_{\tiny\yng(4)}(Z),\qquad 
  \chi_{\tiny\yng(3)}(Z)\lambda_2 =\chi_{\tiny\yng(3,1)}(Z),$$
which should be true when $\lambda_1\gg\lambda_2$.

\myfig{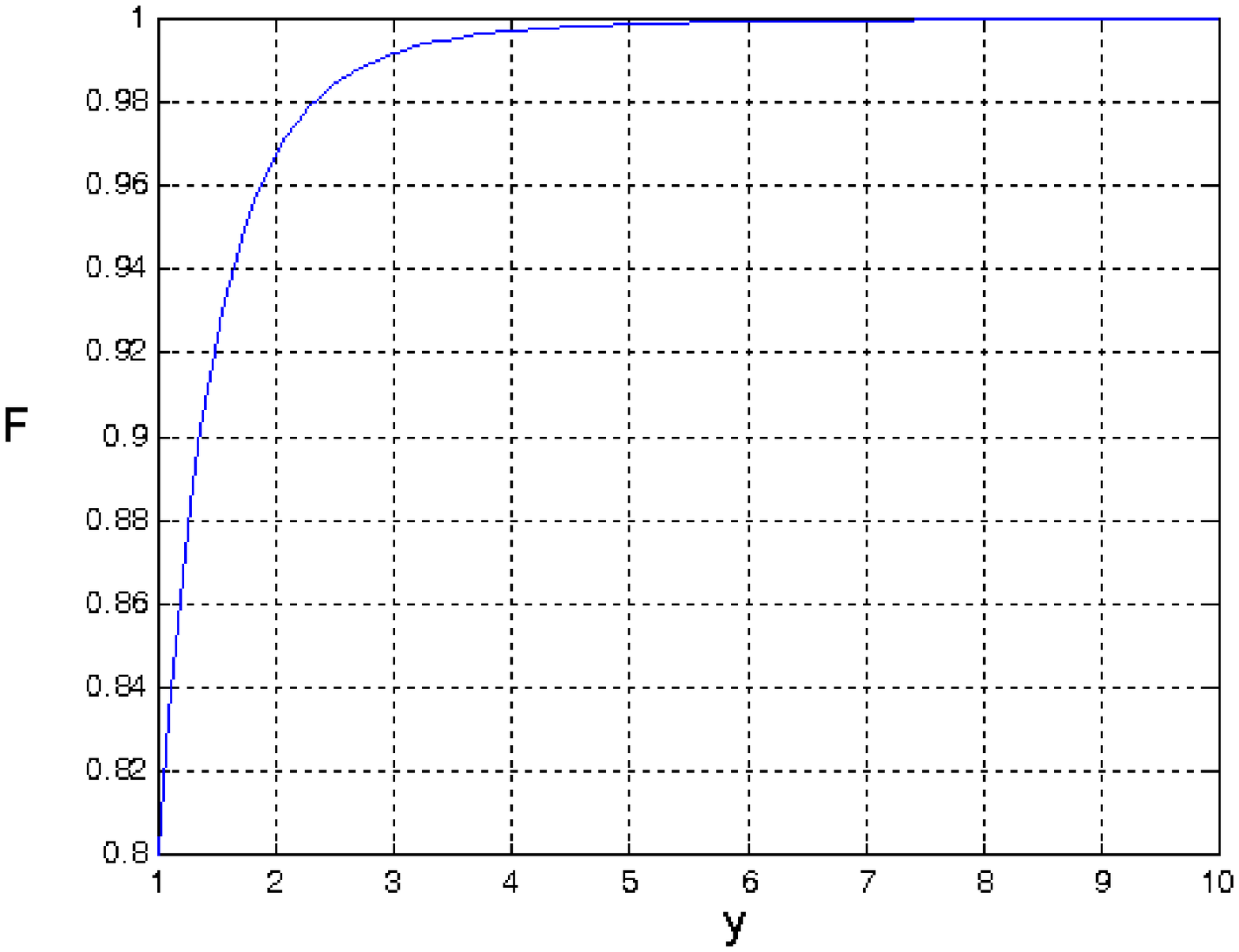}{12.0}{A plot of $F$ against $y=\lambda_1/\lambda_2$.}

In figure 8 we have plotted
$$ F= {\chi_{\tiny\yng(3)}(Z)\lambda_2 \over \chi_{\tiny\yng(3,1)}(Z)},$$
against $y={\lambda_1\over\lambda_2}$. As $y$ increases, $F$ very rapidly approaches 1, confirming our proposal.

Thus, locality on the Young diagram maps, at least for widely separated eigenvalues, into locality on the eigenvalues.  

\section{Beyond LLM}

The LLM geometries correspond, in the super Yang-Mills theory, to operators built using a single matrix $Z$. In this section we would 
like to consider geometries which correspond to operators built using two matrices $X$ and $Z$. These backgrounds will include the 
${1\over 4}$ BPS geometries of \cite{Lunin:2008tf}; we are not yet able to make contact with these geometries. The operators we have in mind are the
restricted Schur polynomials of \cite{Bhattacharyya:2008rb}
\begin{equation}
\chi_{R,(r_n,r_m)}(Z,X)={1\over n! m!}\sum_{\sigma\in S_{n+m}}\Tr_{(r_n,r_m)}(\Gamma_R(\sigma))Z^{i_1}_{i_{\sigma (1)}}\cdots
Z^{i_n}_{i_{\sigma (n)}}X^{i_{n+1}}_{i_{\sigma (n+1)}}\cdots X^{i_{n+m}}_{i_{\sigma (n+m)}}.
\label{restschur}
\end{equation}
The two point function of these operators is trivial to compute using the results of \cite{Bhattacharyya:2008rb}. They also satisfy
a product rule as shown in \cite{Bhattacharyya:2008xy}. Thus we can compute the correlators needed to probe the geometry with gravitons.
The precise form of the product rule for restricted Schur polynomials involves the computation of restricted traces; we do not yet
have efficient methods for this. However, in the special case that $r_n$ is a Young diagram with $M$ columns
and each column contains exactly $N$ boxes we can write\cite{Bhattacharyya:2008xy}
$$ \chi_{+r_m,(r_n,r_m)}(Z,X)={({\rm hooks})_{r_n}({\rm hooks})_{r_m}\over ({\rm hooks})_{+r_m}}
 \chi_{r_n}(Z) \chi_{r_m}(X).$$
After normalizing this becomes (the hat denotes a normalized operator)
$$ \hat{\chi}_{+r_m,(r_n,r_m)}(Z,X)= {\chi_{r_n}(Z)\over \sqrt{f_{r_n}} } {\chi_{r_m}(X)\over \sqrt{f_{r_m}} }\equiv \chi_B(Z,X).$$
Clearly, in this case the required product rule is just the Litttlewood-Richardson rule. Thus, we can study these backgrounds using
the methods of section 3. We will not be so general: in what follows we will assume that $r_n$ and $r_m$ 
are both rectangles that have $N$ rows and that
$r_n$ has $M_n$ columns and $r_m$ $M_m$ columns. We take both $M_n$ and $M_m$ to be $O(N)$ so that $\mu_n\equiv {M_n\over N}$ and
$\mu_m\equiv {M_m\over N}$ are numbers of $O(1)$. The backgrounds we consider are perhaps the most trivial that could be considered. Indeed, 
all traces that appear contain only $X$s or $Z$s. It would be much more interesting to consider backgrounds which are dual to operators 
that have traces over a product of $X$s and $Z$s. 

It is straight forward to verify that
$$ {\cal O}_p ={1\over (1+\mu_n)^{p/2}\sqrt{p!}}\left({\chi_{\tiny \yng(1)}(Z)\over\sqrt{N}}\right)^p \chi_B(Z,X), $$
$$ {\cal O}_p ={1\over (1+\mu_m)^{p/2}\sqrt{p!}}\left({\chi_{\tiny \yng(1)}(X)\over\sqrt{N}}\right)^p \chi_B(Z,X), $$
are normalized $p$ graviton states, as are
$$ {\cal O}_{-p} ={1\over \mu_n^{p/2}\sqrt{p!}}\left({\Tr \left({d\over dZ}\right)\over\sqrt{N}}\right)^p \chi_B(Z,X) ,$$
$$ {\cal O}_{-p} ={1\over \mu_m^{p/2}\sqrt{p!}}\left({\Tr \left({d\over dX}\right)\over\sqrt{N}}\right)^p \chi_B(Z,X) .$$
Thus, even in this more general two matrix background we can define operators dual to local gravitons.
$\Tr(X)$ and $\Tr (Z)$ create these gravitons and the state must again be normalized by the factor $c/N$ that appeared above;
this suggests that these gravitons propagate on some classical geometry. For the gravitons created by $\Tr (Z)$ $c$ is read
from $r_n$; for the gravitons created by $\Tr (X)$ $c$ is read from $r_m$.
This result is easy to interpret: the gravitons built using $Z$s feel only the background of $\chi_{r_n}(Z)$ whilst the
gravitons built using $X$s feel only the background of $\chi_{r_m}(X)$. This is a consequence of the fact that our background
takes the form of a Schur polynomial in $X$ times a Schur polynomial in $Z$. For more general backgrounds we do not expect that
this will be the case. This result suggests that it may be possible to construct some nontrivial ${1\over 4}$ BPS geometries by
cleverly combining LLM geometries.

\section{Discussion}

The Schur polynomials (and their multimatrix generalizations) provide a convenient way 
to organize the interactions of a matrix model - by using the Schur polynomials we have
been able to compute correlators that involve $O(N^2)$ fields. Direct confrontation of
these correlators using standard Feynman diagrams is daunting; using the Schur polynomials,
these computations are simple.
The Schur polynomials have two properties that are ultimately responsible for this
simplicity: (i) one can compute their two point function exactly and further they
diagonalize the two point function; (ii) they satisfy a simple product rule which allows
a straight forward computation of $n$-point correlators.

The Schur polynomials are a nontrivial linear combination of terms which each have a 
different trace structure - the number of traces and the number of fields in each trace
vary from term to term. The Young diagram is a book keeping device which keeps track of 
this structure i.e. it summarizes exactly how the Schur polynomial is constructed.

We have demonstrated some features of how the {\it local} geometry dual to operators with a large 
${\cal R}$-charge emerges from the super Yang-Mills theory. 
In particular, we have described how to define super Yang-Mills operators that are dual to localized
excitations in the bulk of the dual quantum gravity.
The Young diagram labeling of 
the operators has played a crucial role. A Young diagram can be translated into an LLM boundary
condition - it corresponds to a set of concentric rings. Any LLM boundary condition that is
a set of concentric rings can be translated into a Young diagram. Localized excitations in spacetime are 
also localized on the Young diagram. An excitation at a specific radius in the LLM plane has
an energy equal to the radius squared; thus by constructing excitations of a specific energy we
have constructed excitations localized at a specific radius and by measuring the energy of the
excitation we can determine where the excitation is localized. We have defined local creation and annihilation operators 
that add or remove boxes at a specific location on the Young diagram. These operators are 
normalized with a factor equal to ${c\over N}$ where $c$ is the weight of the box which
was added/removed. The excitations that we have studied are confined to move on the $y=0$ LLM
plane and orbit with $\dot{\phi}=1$ at a fixed radius $r_o$. This normalization factor ${c\over N}$ 
where $c$ is the weight of the box corresponding to the localized excitation, is also equal to the 
radius squared $r_o^2$. Of course, as we have reviewed, the weights of the Young diagram encode 
the specific way in which the indices of the matrices making up the operator were symmetrized; we see
that they also tell us the radii of the rings of the LLM boundary condition. 

The Hamiltonian obtained from the field theory measures energy with respect to the background.
By introducing a rescaled time coordinate $\tau=ct$, we can ensure that the Hamiltonian measures
energy with respect to the true vacuum of the theory. In this way, we are lead rather naturally to 
rescaling time differently for the various sectors of the field theory. The LLM geometries have a time 
coordinate that is warped, so that there are various ``local'' times as well as the global time coordinate. 
Our work might provide a natural way to understand these time rescalings in the field theory.

A standard way to describe the large $N$ limit of a matrix model is in terms of a density of eigenvalues.
A particular background would correspond to a specific eigenvalue density. In this article we have used
a Young diagram to specify the background. We have seen however, that this Young diagram can be translated
into a particular eigenvalue distribution. A Young diagram describing $m$ rings will correspond to a
distribution in which the eigenvalues are clustered into $m$ ``clumps'' with eigenvalues in a particular
clump having similar magnitudes. This suggests that the free fermion picture of the eigenvalue dynamics has 
a potential with $m$ minima. The locality that we have seen emerge from the field theory emerges when the 
clumps of eigenvalues have well separated values. This matches nicely with other recent results\cite{Berenstein}.

Finally, note that the reorganization of the degrees of freedom that we have discussed in this article has
a natural extension to multimatrix models. Indeed, the multimatrix operators given in \cite{Bhattacharyya:2008rb} have 
a diagonal two point function and satisfy a generalization of the Littlewood-Richardson rule\cite{Bhattacharyya:2008xy}. 
More work is however needed to write the product rule satisfied by the restricted Schur polynomials in a useful form.
See also \cite{Brown:2007xh},\cite{Kimura:2007wy} for alternative multimatrix operators that can also be analyzed by exploiting
group theory techniques.

{\vskip 0.25cm}

\noindent
{\it Acknowledgements:}  
I would like to thank Norman Ives, Jeff Murugan, Joao Rodrigues, Michael Stephanou and Alex Welte for enjoyable, 
helpful discussions. This work is based upon research supported by the South African Research Chairs Initiative 
of the Department of Science and Technology and National Research Foundation. Any opinion, findings and conclusions 
or recommendations expressed in this material are those of the authors and therefore the NRF and DST do not accept 
any liability with regard thereto. This work is also supported by NRF grant number Gun 2047219.

\appendix

\section{A Review of the LLM Geometries}

The LLM geometries\cite{Lin:2004nb} are regular ${1\over 2}$ BPS solutions of type IIB string theory that are asymptotically AdS$_5\times$S$^5$.
They have an $R\times SO(4)\times SO(4)$ isometry group. The metric is given by
$$ ds^2= -h^{-2}(dt+V_i dx^i)^2 +h^2 (dy^2 +dx^idx^i)+ye^G d\Omega_3^2 + ye^{-G}d\tilde{\Omega}_3^2 \, ,$$
where
$$ h^{-2}=2y\cosh G,\qquad z={1\over 2}\tanh G,$$
$$ y\partial_y V_i=\epsilon_{ij}\partial_j z,\qquad y(\partial_i V_j-\partial_j V_i)=\epsilon_{ij}\partial_y z .$$
The function $z$ is obtained by solving
$$ \partial_i\partial_i z+y\partial_y {\partial_y z\over y}=0 .$$
Regularity requires that $z=\pm {1\over 2}$ on the $y=0$ plane. We will often trade the $x^1,x^2$ coordinates of the $y=0$
plane for an angle and a radius $r,\phi$ or for the complex coordinates $z=x^1+ix^2$, $\bar{z}=x^1-ix^2$.

For a set of rings having a total of $E$ edges with radii $R_l$ $l=1,2,...,E$ we have
$$ z = \sum_{l=1}^E {(-1)^{E-l}\over 2}\left( {r^2+y^2-R_l^2\over\sqrt{(r^2+y^2+R_l^2)^2-4r^2 R_l^2}}-1\right),$$
$$ V_{\phi}(x^1,x^2,y)=\sum_{l=1}^E {(-1)^{E-l+1}\over 2}\left( {r^2+y^2+R_l^2\over\sqrt{(r^2+y^2+R_l^2)^2-4r^2 R_l^2}}-1\right).$$

The fast string limit\cite{Kruczenski:2003gt} of the Polyakov action for a string 
moving in an LLM geometry is given by\cite{Vazquez:2006id,Chen:2007gh}
$$ S=L\int d\tau\int_0^1 d\sigma\left( {i\over 2}V\dot{\bar{z}}-{i\over 2}\bar{V}\dot{z}-{\lambda\over L^2} \partial_\sigma z
\partial_\sigma\bar{z}\right) .$$

\section{Correlators}

In this section we will compute the correlators needed to probe a general background $B$ with a closed string. In terms of the loop
$${\cal O}=\Tr (YZ^{n_1}_a YZ^{n_2}_a Y\cdots YZ^{n_L}_a), $$
we wish to compute
$$\left\langle {\cal O}{\cal O}^\dagger \right\rangle_B\equiv \left\langle  {\chi_B\over\sqrt{f_B}}{\cal O}{\cal O}^\dagger {\chi_B^\dagger\over\sqrt{f_B}} \right\rangle .$$
Assume the $n_i$ are all distinct. The leading contribution at large $N$ comes from the following contraction of the $Y$s
$$\left\langle {\cal O}{\cal O}^\dagger \right\rangle_B\equiv \left\langle  {\chi_B\over\sqrt{f_B}}
\prod_{i=1}^L \Tr (Z_a^{n_i}(Z_a^\dagger)^{n_i}) {\chi_B^\dagger\over\sqrt{f_B}} \right\rangle .$$

Look back at the correlation functions (\ref{gravitoncorr}) and (\ref{gravitoncorrtwo}). These correlators are easy to reproduce
using the two point function
$$\left\langle Z_{ij}Z^\dagger_{kl}\right\rangle =\delta_{il}\delta_{jk} .$$
Now consider (\ref{corr1}) and (\ref{multimom}). These correlators are clearly reproduced if we take
$$\left\langle (Z_a)_{ij}(Z_a)^\dagger_{kl}\right\rangle =(1+\mu)\delta_{il}\delta_{jk} .$$
By considering more general correlators, it is straight forward to see that all such correlators are reproduced
by the rule
\begin{equation}
\left\langle (Z_{\rm loc})_{ij}(Z_{\rm loc})^\dagger_{kl}\right\rangle ={c\over N}\delta_{il}\delta_{jk} ,
\label{ribbon}
\end{equation}
where $Z_{\rm loc}$ acts locally on the Young diagram, adding boxes with weight $c$ up to addition of an irrelevant (at large $N$)
$O(1)$ number. This clearly reproduces the correlators computed using Schur technology.

Focus again on the case that $B$ is the annulus. With the result (\ref{ribbon}) in hand we find
$$ \left\langle {\cal O}{\cal O}^\dagger \right\rangle_B=
\left\langle  {\chi_B\over\sqrt{f_B}} \prod_{i=1}^L \Tr (Z_a^{n_i}(Z_a^\dagger)^{n_i}) {\chi_B^\dagger\over\sqrt{f_B}} \right\rangle =
\prod_{i=1}^L N^{n_i+1}(1+\mu)^{n_i}. $$
Clearly it is now straight forward to compute correlators needed when probing the background with a closed string state.

A very convenient way to account for the planar contractions is by introducing Cuntz oscillators\cite{Gopakumar:1994iq}. The matrix two point function 
determines the Cuntz oscillator algebra. In terms of Cuntz oscillators (\ref{ribbon}) becomes
$$\alpha\alpha^\dagger ={c\over N},\qquad \alpha^\dagger\alpha ={c\over N}(1-|0\rangle\langle 0|).$$
The analysis of \cite{Berenstein:2006qk} can now be applied to obtain the Cuntz Hamiltonian quoted in section 3.1.


\begin{thebibliography}{30}
\parskip-2pt

\bibitem{Maldacena:1997re}
  J.~M.~Maldacena,
  ``The large N limit of superconformal field theories and supergravity,''
  Adv.\ Theor.\ Math.\ Phys.\  {\bf 2}, 231 (1998)
  [Int.\ J.\ Theor.\ Phys.\  {\bf 38}, 1113 (1999)]
  [arXiv:hep-th/9711200];\\
  S.~S.~Gubser, I.~R.~Klebanov and A.~M.~Polyakov,
  ``Gauge theory correlators from non-critical string theory,''
  Phys.\ Lett.\ B {\bf 428}, 105 (1998)
  [arXiv:hep-th/9802109];\\
  E.~Witten,
  ``Anti-de Sitter space and holography,''
  Adv.\ Theor.\ Math.\ Phys.\  {\bf 2}, 253 (1998)
  [arXiv:hep-th/9802150].

\bibitem{Lin:2004nb}
  H.~Lin, O.~Lunin and J.~M.~Maldacena,
  ``Bubbling AdS space and 1/2 BPS geometries,''
  JHEP {\bf 0410}, 025 (2004)
  [arXiv:hep-th/0409174].

\bibitem{Skenderis:2007yb}
L.~F.~Alday, J.~de Boer and I.~Messamah,
  ``What is the dual of a dipole?,''
  Nucl.\ Phys.\  B {\bf 746}, 29 (2006)
  [arXiv:hep-th/0511246],\\
  V.~Balasubramanian, B.~Czech, K.~Larjo and J.~Simon,
  ``Integrability vs. information loss: A simple example,''
  JHEP {\bf 0611}, 001 (2006)
  [arXiv:hep-th/0602263],\\
  V.~Balasubramanian, B.~Czech, K.~Larjo, D.~Marolf and J.~Simon,
  ``Quantum geometry and gravitational entropy,''
  JHEP {\bf 0712}, 067 (2007)
  [arXiv:0705.4431 [hep-th]],\\
  K.~Skenderis and M.~Taylor,
  ``Anatomy of bubbling solutions,''
  JHEP {\bf 0709}, 019 (2007)
  [arXiv:0706.0216 [hep-th]].

\bibitem{Lee:1998bxa}
  S.~Lee, S.~Minwalla, M.~Rangamani and N.~Seiberg,
   ``Three-point functions of chiral operators in D = 4, N = 4 SYM at  large N,''
  Adv.\ Theor.\ Math.\ Phys.\  {\bf 2}, 697 (1998)
  [arXiv:hep-th/9806074],\\
  K.~A.~Intriligator,
   ``Bonus symmetries of N = 4 super-Yang-Mills correlation functions via  AdS
  duality,''
  Nucl.\ Phys.\  B {\bf 551}, 575 (1999)
  [arXiv:hep-th/9811047],\\
  B.~U.~Eden, P.~S.~Howe, A.~Pickering, E.~Sokatchev and P.~C.~West,
  ``Four-point functions in N = 2 superconformal field theories,''
  Nucl.\ Phys.\  B {\bf 581}, 523 (2000)
  [arXiv:hep-th/0001138],\\
  B.~U.~Eden, P.~S.~Howe, E.~Sokatchev and P.~C.~West,
   ``Extremal and next-to-extremal n-point correlators in four-dimensional
     SCFT,''
  Phys.\ Lett.\  B {\bf 494}, 141 (2000)
  [arXiv:hep-th/0004102].

\bibitem{Corley:2001zk}
  S.~Corley, A.~Jevicki and S.~Ramgoolam,
  ``Exact correlators of giant gravitons from dual N = 4 SYM theory,''
  Adv.\ Theor.\ Math.\ Phys.\  {\bf 5}, 809 (2002)
  [arXiv:hep-th/0111222].

\bibitem{Berenstein:2004kk}
  D.~Berenstein,
  ``A toy model for the AdS/CFT correspondence,''
  JHEP {\bf 0407}, 018 (2004)
  [arXiv:hep-th/0403110].

\bibitem{Brown:2007xh}
  T.~W.~Brown, P.~J.~Heslop and S.~Ramgoolam,
  ``Diagonal multi-matrix correlators and BPS operators in N=4 SYM,''
  arXiv:0711.0176 [hep-th].

\bibitem{Kimura:2007wy}
  Y.~Kimura and S.~Ramgoolam,
  ``Branes, Anti-Branes and Brauer Algebras in Gauge-Gravity duality,''
  arXiv:0709.2158 [hep-th].

\bibitem{Bhattacharyya:2008rb}
  R.~Bhattacharyya, S.~Collins and R.~de~Mello~Koch,
  ``Exact Multi-Matrix Correlators,''
  arXiv:0801.2061 [hep-th].

\bibitem{sanjaye}
S.~Ramgoolam,
``Schur-Weyl duality as an instrument of Gauge-String duality,''
  arXiv:0804.2764 [hep-th].

\bibitem{Vazquez:2006id}
  S.~E.~Vazquez,
  ``Reconstructing 1/2 BPS space-time metrics from matrix models and spin
  chains,''
  Phys.\ Rev.\  D {\bf 75}, 125012 (2007)
  [arXiv:hep-th/0612014].

\bibitem{Chen:2007gh}
  H.~Y.~Chen, D.~H.~Correa and G.~A.~Silva,
  ``Geometry and topology of bubble solutions from gauge theory,''
  Phys.\ Rev.\  D {\bf 76}, 026003 (2007)
  [arXiv:hep-th/0703068].

\bibitem{Brezin:1977sv}
  E.~Brezin, C.~Itzykson, G.~Parisi and J.~B.~Zuber,
  ``Planar Diagrams,''
  Commun.\ Math.\ Phys.\  {\bf 59}, 35 (1978).

\bibitem{McGreevy:2000cw}
  J.~McGreevy, L.~Susskind and N.~Toumbas,
  ``Invasion of the giant gravitons from anti-de Sitter space,''
  JHEP {\bf 0006}, 008 (2000)
  [arXiv:hep-th/0003075].

\bibitem{Corley:2002mj}
  S.~Corley and S.~Ramgoolam,
  ``Finite factorization equations and sum rules for BPS correlators in  N = 4
  SYM theory,''
  Nucl.\ Phys.\ B {\bf 641}, 131 (2002)
  [arXiv:hep-th/0205221].

\bibitem{related}
  R.~de Mello Koch and R.~Gwyn,
  ``Giant graviton correlators from dual SU(N) super Yang-Mills theory,''
  JHEP {\bf 0411}, 081 (2004)
  [arXiv:hep-th/0410236].

\bibitem{de Mello Koch:2007uu}
  R.~de Mello Koch, J.~Smolic and M.~Smolic,
  ``Giant Gravitons - with Strings Attached (I),'' JHEP {\bf 0706}, 074 (2007),
  arXiv:hep-th/0701066.

\bibitem{Mandal:2005wv}
  G.~Mandal,
  ``Fermions from half-BPS supergravity,''
  JHEP {\bf 0508}, 052 (2005)
  [arXiv:hep-th/0502104].

\bibitem{Balasubramanian:2004nb}
  V.~Balasubramanian, D.~Berenstein, B.~Feng and M.~x.~Huang,
  ``D-branes in Yang-Mills theory and emergent gauge symmetry,''
  JHEP {\bf 0503}, 006 (2005)
  [arXiv:hep-th/0411205].

\bibitem{de Mello Koch:2007uv}
  R.~de Mello Koch, J.~Smolic and M.~Smolic,
  ``Giant Gravitons - with Strings Attached (II),'' JHEP {\bf 0709} 049 (2007),
  arXiv:hep-th/0701067.

\bibitem{Balasubramanian:2005mg}
  V.~Balasubramanian, J.~de Boer, V.~Jejjala and J.~Simon,
  ``The library of Babel: On the origin of gravitational thermodynamics,''
  JHEP {\bf 0512}, 006 (2005)
  [arXiv:hep-th/0508023],\\
  V.~Balasubramanian, V.~Jejjala and J.~Simon,
  ``The library of Babel,''
  Int.\ J.\ Mod.\ Phys.\  D {\bf 14}, 2181 (2005)
  [arXiv:hep-th/0505123].

\bibitem{Takayama:2005yq}
  Y.~Takayama and A.~Tsuchiya,
  ``Complex matrix model and fermion phase space for bubbling AdS
    geometries,''
  JHEP {\bf 0510}, 004 (2005)
  [arXiv:hep-th/0507070].

\bibitem{Lunin:2008tf}
A.~Donos,
  ``A description of 1/4 BPS configurations in minimal type IIB SUGRA,''
  Phys.\ Rev.\  D {\bf 75}, 025010 (2007)
  [arXiv:hep-th/0606199],\\
  B.~Chen {\it et al.},
  ``Bubbling AdS and droplet descriptions of BPS geometries in IIB
  supergravity,''
  JHEP {\bf 0710}, 003 (2007)
  [arXiv:0704.2233 [hep-th]],\\
  O.~Lunin,
  ``Brane webs and 1/4-BPS geometries,''
  arXiv:0802.0735 [hep-th].

\bibitem{Bhattacharyya:2008xy}
  R.~Bhattacharyya, R.~de~Mello~Koch and M.~Stephanou,
  ``Exact Multi-Restricted Schur Polynomial Correlators,''
  arXiv:0805.3025 [hep-th].

\bibitem{Berenstein}
  D.~Berenstein, M.~Hanada and S.~A.~Hartnoll,
  ``Multimatrix-Models and Emergent Geometry,''
  arXiv:0805.4658 [hep-th],\\
  D.~Berenstein,
  ``A strong coupling expansion for N=4 SYM theory and other SCFT's,''
  arXiv:0804.0383 [hep-th],\\
  D.~E.~Berenstein and S.~A.~Hartnoll,
  ``Strings on conifolds from strong coupling dynamics: quantitative results,''
  JHEP {\bf 0803} (2008) 072
  [arXiv:0711.3026 [hep-th]],\\
  D.~Berenstein,
  ``Strings on conifolds from strong coupling dynamics, part I,''
  JHEP {\bf 0804} (2008) 002
  [arXiv:0710.2086 [hep-th]],\\
  D.~Berenstein,
  ``Large N BPS states and emergent quantum gravity,''
  JHEP {\bf 0601} (2006) 125
  [arXiv:hep-th/0507203].

\bibitem{Kruczenski:2003gt}
  M.~Kruczenski,
  ``Spin chains and string theory,''
  Phys.\ Rev.\ Lett.\  {\bf 93}, 161602 (2004)
  [arXiv:hep-th/0311203].
  M.~Kruczenski, A.~V.~Ryzhov and A.~A.~Tseytlin,
   ``Large spin limit of AdS(5) x S**5 string theory and low energy  expansion
     of ferromagnetic spin chains,''
  Nucl.\ Phys.\  B {\bf 692}, 3 (2004)
  [arXiv:hep-th/0403120].

\bibitem{Berenstein:2006qk}
  D.~Berenstein, D.~H.~Correa and S.~E.~Vazquez,
  ``A study of open strings ending on giant gravitons, spin chains and
  integrability,''
  [arXiv:hep-th/0604123],\\
  D.~Berenstein, D.~H.~Correa and S.~E.~Vazquez,
  ``Quantizing open spin chains with variable length: An example from giant
    gravitons,''
  Phys.\ Rev.\ Lett.\  {\bf 95}, 191601 (2005)
  [arXiv:hep-th/0502172],\\
  D.~H.~Correa and G.~A.~Silva,
  ``Dilatation operator and the super Yang-Mills duals of open strings on AdS
    giant gravitons,''
  JHEP {\bf 0611}, 059 (2006)
  [arXiv:hep-th/0608128].

\bibitem{Gopakumar:1994iq}
  R.~Gopakumar and D.~J.~Gross,
  ``Mastering the master field,''
  Nucl.\ Phys.\  B {\bf 451}, 379 (1995)
  [arXiv:hep-th/9411021].


\end{thebibliography}
\end{document}